\documentclass[structabstract]{aa}  
\usepackage{graphicx}
\usepackage{times}
\usepackage{subfigure}
\usepackage{natbib}
\usepackage{txfonts}
\newcommand{\beq}{\begin{equation}}
\newcommand{\eeq}{\end{equation}}

\def\gs{\mathrel{\lower0.6ex\hbox{$\buildrel {\textstyle >}\over{\scriptstyle \sim}$}}}
\def\ls{\mathrel{\lower0.6ex\hbox{$\buildrel {\textstyle <}\over{\scriptstyle \sim}$}}}

\begin{document}
   \title{A multi-wavelength strong lensing analysis of baryons and dark matter in the dynamically active cluster AC~114}

%   \subtitle{I. Overviewing the $\kappa$-mechanism}

   \author{M. Sereno
          \inst{1,2,3}\fnmsep\thanks{\email{mauro.sereno@polito.it}}
          \and
          M. Lubini\inst{1}
          \and
          Ph. Jetzer\inst{1}
          }

   \institute{
   Institut f\"{u}r Theoretische Physik, Universit\"{a}t Z\"{u}rich, Winterthurerstrasse 190, 8057 Z\"{u}rich, Switzerland    
 \and   
 Dipartimento di Fisica, Politecnico di Torino, Corso Duca degli Abruzzi 24, 10129 Torino, Italia
 \and
INFN, Sezione di Torino, Via Pietro Giuria 1, 10125, Torino, Italia       
}

%   \date{Received September 15, 1996; accepted March 16, 1997}
%   \date{19 April 2009}

% \abstract{}{}{}{}{} 
% 5 {} token are mandatory
 
  \abstract
   {Strong lensing studies can provide detailed mass maps of the inner regions even in dynamically active galaxy clusters. 
   }
   {It is shown that proper modelling of the intracluster medium, i.e. the main baryonic component, can play an important role. In fact, the addition of a new contribution accounting for the gas can increase the statistical significance of the lensing model.}
   {We propose a parametric method for strong lensing analyses which exploits multi-wavelength observations. The mass model accounts for cluster-sized dark matter halos, galaxies (whose stellar mass can be obtained from optical analyses) and the intracluster medium. The gas distribution is fitted to lensing data exploiting prior knowledge from X-ray observations. This gives an unbiased look at each matter component and allows us to study the dynamical status of a cluster.}
   {The method has been applied to AC~114, an irregular X-ray cluster. We find positive evidence for dynamical activity, with the dark matter distribution shifted and rotated with respect to the gas. On the other hand, the dark matter follows the galaxy density both for shape and orientation, which hints at its collisionless nature. The inner region ($\ls 250~\mathrm{kpc}$) is under-luminous in optical bands whereas the gas fraction ($\sim 20 \pm 5 \%$) slightly exceeds typical values. Evidence from lensing and X-ray suggests that the cluster develops in the plane of the sky and is not affected by the lensing over-concentration bias. Despite the dynamical activity, the matter distribution seems to be in agreement with predictions from $N$-body simulations. An universal cusped profile provides a good description of either the overall or the dark matter distribution whereas theoretical scaling relations seem to be nicely fitted.
% the inner slope of the global density profile $\rho \sim r^{-\alpha}$, as inferred from a power-law analysis, seems to be slightly steeper ($\alpha \sim 1.3$) than the NFW prediction. The total matter distribution has concentration $c_{200} = 3.5 \pm 0.7$ and mass $M_{200}=(1.3\pm0.9)\times 10^{15}M_\odot$, which fit nicely theoretical scaling relations.
}
   {}


\titlerunning{Multi-wavelength lensing in AC~114}
   \maketitle
%
%________________________________________________________________

\section{Introduction}

Understanding the formation and evolution of galaxy clusters is an open problem in modern astronomy. On the theoretical side, $N$-body simulations are now able to make detailed statistical predictions on dark matter (DM) halo properties \citep{nav+al97,bul+al01,die+al04,duf+al08}. On the observational side, multi-wavelength observations from the radio to the optical bands to X-ray observations of galaxy clusters can provide a deep insight on real features \citep{clo+al04,def+al05,smi+al05,hic+al06}. Results are impressive on both sides, but further work is still required. Large numerical simulations can not still efficiently embody gas physics, whereas combining multi-wavelength data sets can be misleading if the employed hypotheses (hydrostatic and/or dynamical equilibrium, spherical symmetry, just to list a couple of very common ones) do not hold. So, areas of disagreement between predictions and measurements still persist.

Here, we want to consider a way to exploit multi-wavelength data sets in strong lensing data analyses. Strong lensing modelling can give detailed maps of the inner regions of galaxy clusters without relying on hypotheses on equilibrium and is negligibly affected by projection effects due to large-scale fields or aligned structures. However, massive lensing clusters build up a biased sample for statistical studies \citep{hen+al07,og+bl09}. Multi-wavelength analyses of lensing galaxy clusters have been exploited following different approaches. \citet{smi+al05} compared X-ray and strong lensing maps of intermediate redshift clusters to infer equilibrium criteria. Detailed lensing features can reveal dynamical activity even in apparently relaxed clusters \citep{mir+al08}. Investigations of the bullet cluster showed that dark matter follows the collisionless galaxies whereas the gas is stripped away in mergers \citep{clo+al04}. Comparison of snapshots of active clusters taken with weak lensing, X-ray surface brightness or galaxy luminosity revealed the relative displacement of the different components at different stages of merging \citep{ok+um08}.

The usual way to dissect dark matter from baryons in lensing analyses goes on by first obtaining a map of the total matter distribution fitting the lensing features and then subtracting the gas contribution as inferred from X-ray observations \citep{bra+al08}. The total mass map can be obtained either with parametric models in which the contribution from cluster-sized DM halos can be considered together with the main galactic DM halos \citep{nat+al98,lim+al08} or with non-parametric analyses, where dark matter meso-structures and galactic contributions are seen as deviations from smooth-averaged profiles \citep{sah+al07}. Mass in stars and stellar remnants is estimated from galaxy luminosity assuming suitable stellar mass to light ratios. Such approaches have obvious merits but also some unavoidable shortcomings. 

Collisionless matter and gas are displaced in dynamical active clusters. Furthermore, in relaxed clusters, gas and dark matter profiles have usually different slopes. Since the gas follows the potential, its distribution is usually rounder than dark matter, so that even if the intracluster medium (ICM) and the dark matter are intrinsically aligned their projected masses cast on the sky with different orientation and ellipticity \citep{sta77}. Such features can be missed by usual approaches since the total matter profile may fall short in trying to account for the properties of each component . In fact, the parameters of the model can be not numerous enough to reproduce all the details whereas in non parametric approaches the fitting procedure must be weighted in order to prefer smooth distributions over clumpy ones, which may wash out small scale details such as gas and dark matter off-centred by few arcseconds.

We try to take a step further by exploiting a parametric model which has three kind of components: cluster-sized dark matter halos; galaxy-sized (dark plus stellar) matter halos; cluster-sized gas distribution. In our approach, the ICM distribution is embedded in the strong lensing modelling from the very beginning so avoiding unpleasant biases. In order to reduce the total number of parameters, the X-ray surface brightness data are fully exploited so that the gas contribution is fixed within the observational uncertainties. This allows to constrain the mass model using X-ray data without relying on the assumption of hydrostatic equilibrium. As far as the stellar component is concerned, we take the usual path: total galaxy masses (DM plus baryons) are derived through the lensing fitting procedure whereas the stellar contribution is inferred from luminosity. The main advantage we recognise in such an approach is that we are able to infer directly the dark matter mass. This is the component best (and under some points of view the only one) constrained in numerical simulations so that our novel approach eases the comparison with their theoretical predictions. Furthermore, we will be able to compare the gas distribution with the dark matter, which is an obvious improvement with respect to the usual way of comparing total projected mass distributions with surface brightness maps.

\begin{figure}
        \resizebox{\hsize}{!}{\includegraphics{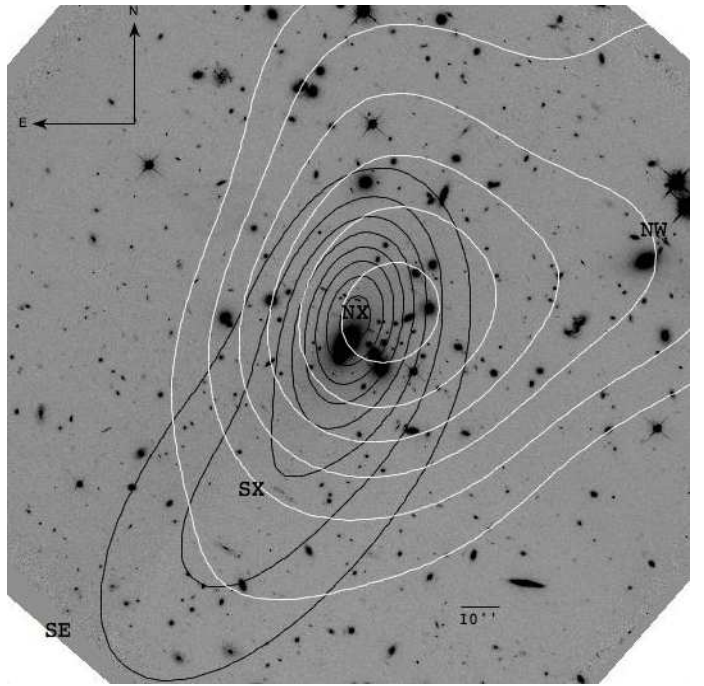}}
        \caption{Grey-scale archive HST/ACS image ($F850$ band) of the core of AC~114. North is up and east is to the left. The field covers $3\arcmin \times 3 \arcmin$. Overlaid are the linearly spaced adaptively-smoothed Chandra surface brightness X-ray inner (black) contours and the inner light (white) contours. Light isophotes have been obtained smoothing the luminosity map with a Gaussian kernel with a dispersion of $30\arcsec$. NX and SX are the centroids of the main X-ray clump and the southern tail, respectively; NW and SE locate likely mass clumps.}
	\label{fig_ac114_iso}
\end{figure}

We apply our method to AC~114. AC~114 is a prototypical example of the Butcher-Oemler effect with a higher fraction of blue, late-type galaxies than in lower redshift clusters, rising to 60\% outside of the core region \citep{cou+al98}. The fraction of interacting galaxies ($\sim 12\%$) is also high \citep{cou+al98}. AC~114 was classified as a Bautz-Morgan type II-III cluster \citep{kr+be07}, suggesting a young dynamical age. The cluster is significantly elongated in the southeast-northwest direction \citep{cou+al01}, see Fig.~\ref{fig_ac114_iso}. There are two main reasons for studying this cluster. First, the core region of AC~114 is rich in multiple images, allowing a very detailed analysis. Redshift contrast between multiple lensed sources can give a good measurement of the enclosed mass at two different radii, thus providing a good estimate of the mass profile in between \citep{sa+re09}. The same kind of information can be obtained also combining strong and weak lensing data, but multiple lensed sources allow us to consistently get the profile slope without mixing systematics from different methods. Furthermore, there are several images very near the cluster centre, which allows to accurately determine the radial slope of the matter distribution in the very inner regions. Despite the abundant data, AC~114 has been object of just a couple of lensing investigations. In the first one from \citet{nat+al98}, later on improved in \citet{cam+al01}, weak lensing constraints were also used and the mass modelling, which was inspired by the optical galaxy distribution, considered a main clump and two additional cluster substructures, see App.~\ref{app_subs}. The second lensing analysis, inspired by the X-ray images, associated each of the two X-ray emitting regions to a dark matter clump in separate hydrostatic equilibrium \citep{def+al04}. Both approaches reproduced the image positions with an accuracy of $\gs 1\arcsec$ but neither one used all of the image systems with confirmed spectroscopic redshifts. So, there is still room for substantial improvement.

Second, AC~114 has significant evidence of experiencing an ongoing merging. We will be able to study the details of the dark matter distribution in a dynamical active cluster, constraining at the same time the properties of the dark matter and the evolution of this interesting cluster. The lensing analysis of the cluster will put us in position to compare estimates with theoretical predictions. 

The paper is as follows. In Sec.~\ref{sec_gala} we discuss the galaxy distribution and portray the luminosity and the number densities. Stellar mass inferred from the measured luminosity is considered as well. Section~\ref{sec_dyna} is devoted to dynamics. We obtain an updated estimate of the cluster mass that will be compared to the lensing results later on. In Sec.~\ref{sec_xray}, we review literature results on the X-ray observations of the cluster and add some new considerations on its dynamical status. Section~\ref{sec_stro} and~\ref{sec_infe}  are devoted to the lensing analysis. In Sec.~\ref{sec_stro}, we review the optical data and the parametric models employed; in Sec.~\ref{sec_infe}, we present our statistical investigation. Section~\ref{sec_resu} lists the results obtained with our multi-wavelength approach, whereas Sec.~\ref{sec_comp} discusses some results in view of theoretical expectations. Section~\ref{sec_disc} is devoted to some final considerations. In App.~\ref{app_vel}, we detail our procedure to estimate the galaxy velocity dispersion $\sigma_\mathrm{los}$. Appendix~\ref{app_subs} is devoted to an analysis of substructures based on classical optical methods. Projection effects are dealt with in App.~\ref{app_proj}.
Throughout the paper, we assume a $\Lambda$CDM cosmology with density parameters $\Omega_\mathrm{M}=0.3$, $\Omega_{\Lambda}=0.7$ and an Hubble constant $H_0=100h~\mathrm{km~s}^{-1}\mathrm{Mpc}^{-1}$, $h=0.7$. This implies a linear scale of $3.22~\mathrm{kpc}/h$ per arcsec at the cluster redshift $z=0.315$. As reference mass and radius for the cluster we will consider $M_{200}$ and $r_{200}$, i.e. the mass and the radius containing an overdensity of 200 times the critical one. We quote uncertainties at the 68.3\% confidence level.

\begin{table*}
\centering
\begin{tabular}[c]{lccccccc}       
\hline        
\noalign{\smallskip}
Density	&$\theta_\mathrm{max}$	& $\beta$	&$\theta_\mathrm{c}$	&$\theta_{1,0}$	&$\theta_{2,0}$	&$\epsilon$	&$\theta_\epsilon$ \\ 
		&($\arcsec$)	& 	&($\arcsec$)	&($\arcsec$)	&($\arcsec$)	&	&($\deg$) \\       
\noalign{\smallskip}
\hline              
\noalign{\smallskip}
Number	&$60$	& $1.2_{-0.4}^{+0.7}$	&$50_{-30}^{+30}$	&$10_{-4}^{+3}$	&$6_{-4}^{+3}$	&$0.25_{-0.11}^{+0.12}$	&$-40_{-20}^{+20}$ \\ 
\noalign{\smallskip}
		&$120$	& $1.07_{-0.15}^{+0.30}$	&$140_{-30}^{+20}$	&$10_{-3}^{+4}$	&$4_{-4}^{+4}$	&$0.53_{-0.04}^{+0.03}$	&$-41_{-3}^{+3}$ \\
\noalign{\smallskip}          
\hline
\noalign{\smallskip}
Luminosity	&$60$	& $1.6_{-0.5}^{+2.3}$	&$36_{-13}^{+50}$	&$7_{-3}^{+3}$	&$7_{-5}^{+3}$	&$0.22_{-0.15}^{+0.10}$	&$-30_{-30}^{+30}$ \\ 
\noalign{\smallskip}
		&$120$	& $0.86_{-0.10}^{+0.60}$	&$42_{-20}^{+70}$	&$13_{-5}^{+5}$	&$11_{-6}^{+6}$	&$0.42_{-0.08}^{+0.06}$	&$-26_{-10}^{+9}$ \\   
\noalign{\smallskip}
\hline
\noalign{\smallskip}
Stellar mass	&$60$	& $1.6_{-0.4}^{+1.5}$	&$36_{-14}^{+30}$	&$8_{-4}^{+3}$	&$6_{-4}^{+3}$	&$0.24_{-0.10}^{+0.10}$	&$-30_{-30}^{+20}$ \\ 
\noalign{\smallskip}
		&$120$	& $0.96_{-0.12}^{+0.17}$	&$36_{-14}^{+20}$	&$10_{-5}^{+5}$	&$9_{-5}^{+5}$	&$0.39_{-0.10}^{+0.07}$	&$-32_{-12}^{+12}$ \\  
\noalign{\smallskip}
\hline 
\end{tabular}       
\centering       
\caption{Properties of the galaxy distributions, modelled as $\beta$ profiles, within circular regions of outer radius $\theta_\mathrm{max}$. Angles are measured North over East.}            
\par\noindent  
\label{tab_gal_den}     
\end{table*}

\section{Galaxy distribution}
\label{sec_gala}

\begin{figure*}
\begin{center}
$\begin{array}{c@{\hspace{.1in}}c@{\hspace{.1in}}c}
\includegraphics[width=7cm]{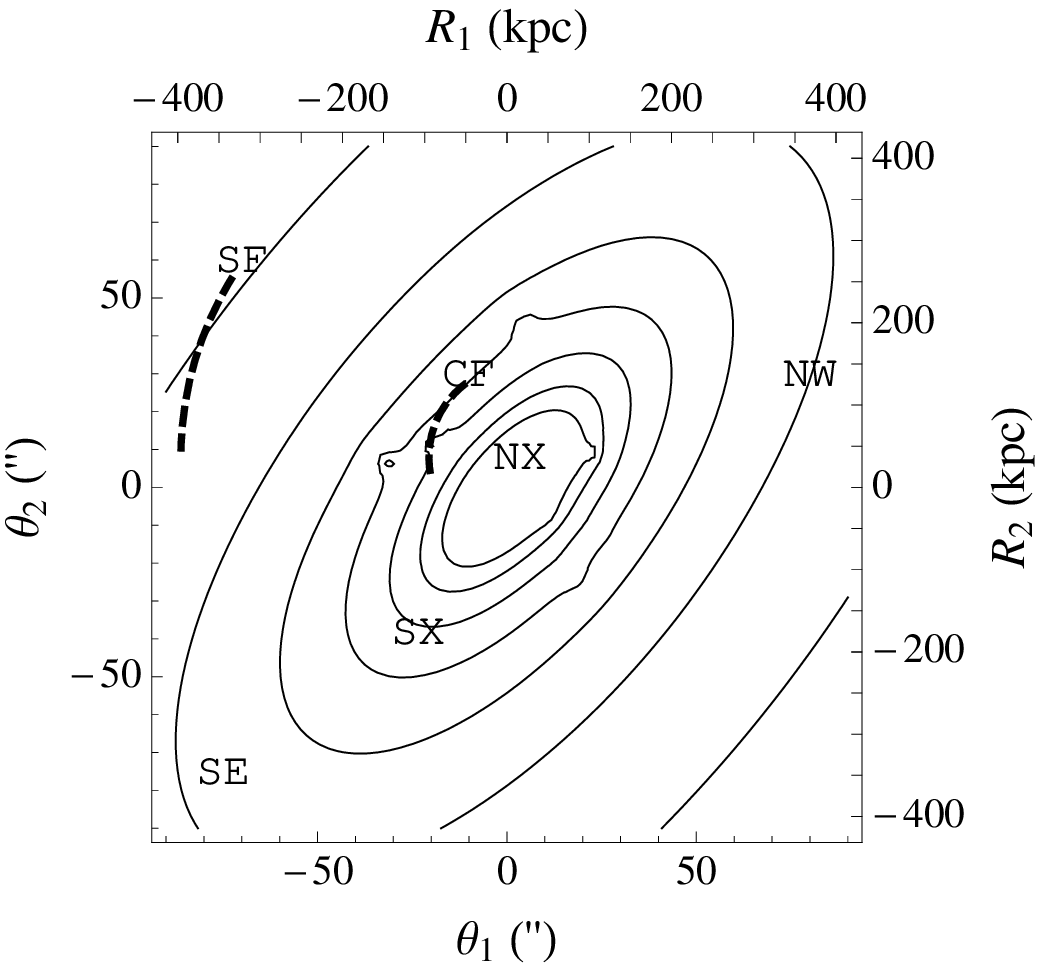} &
\includegraphics[width=7cm]{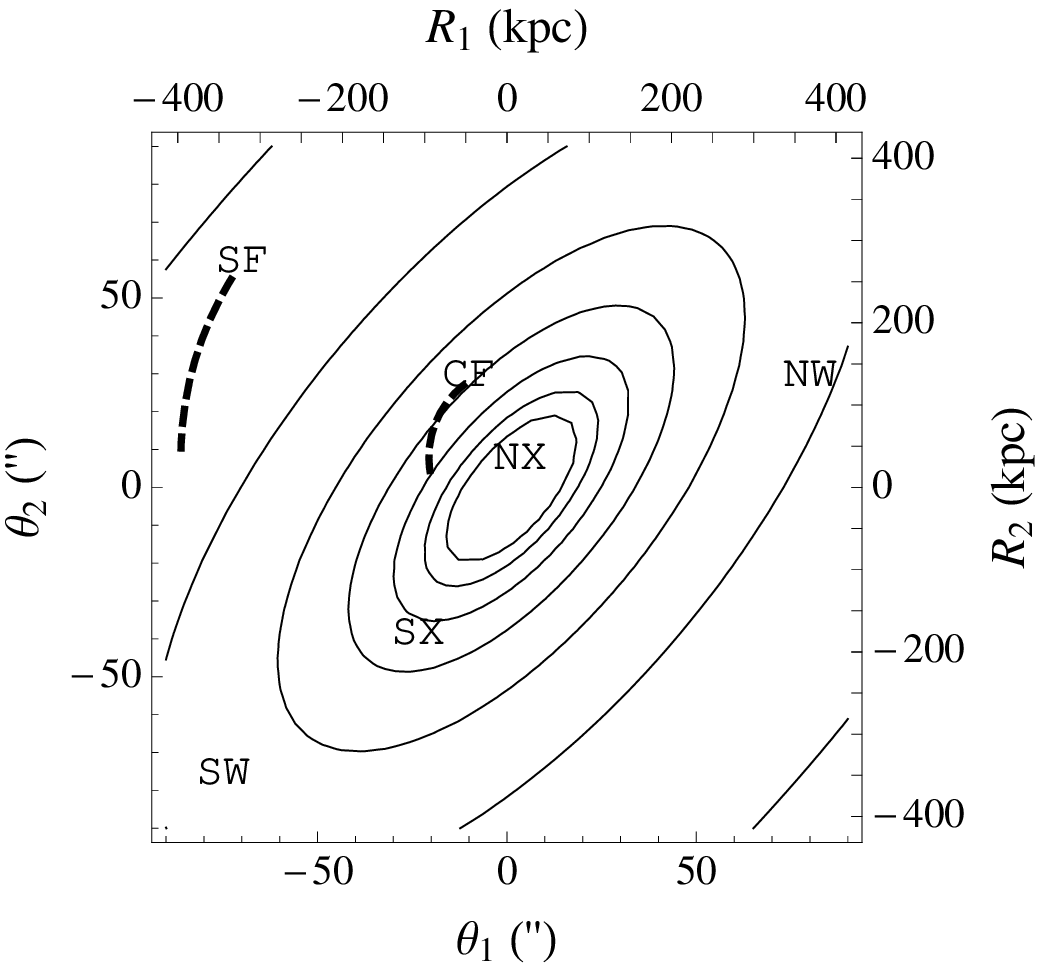} \\
\includegraphics[width=7cm]{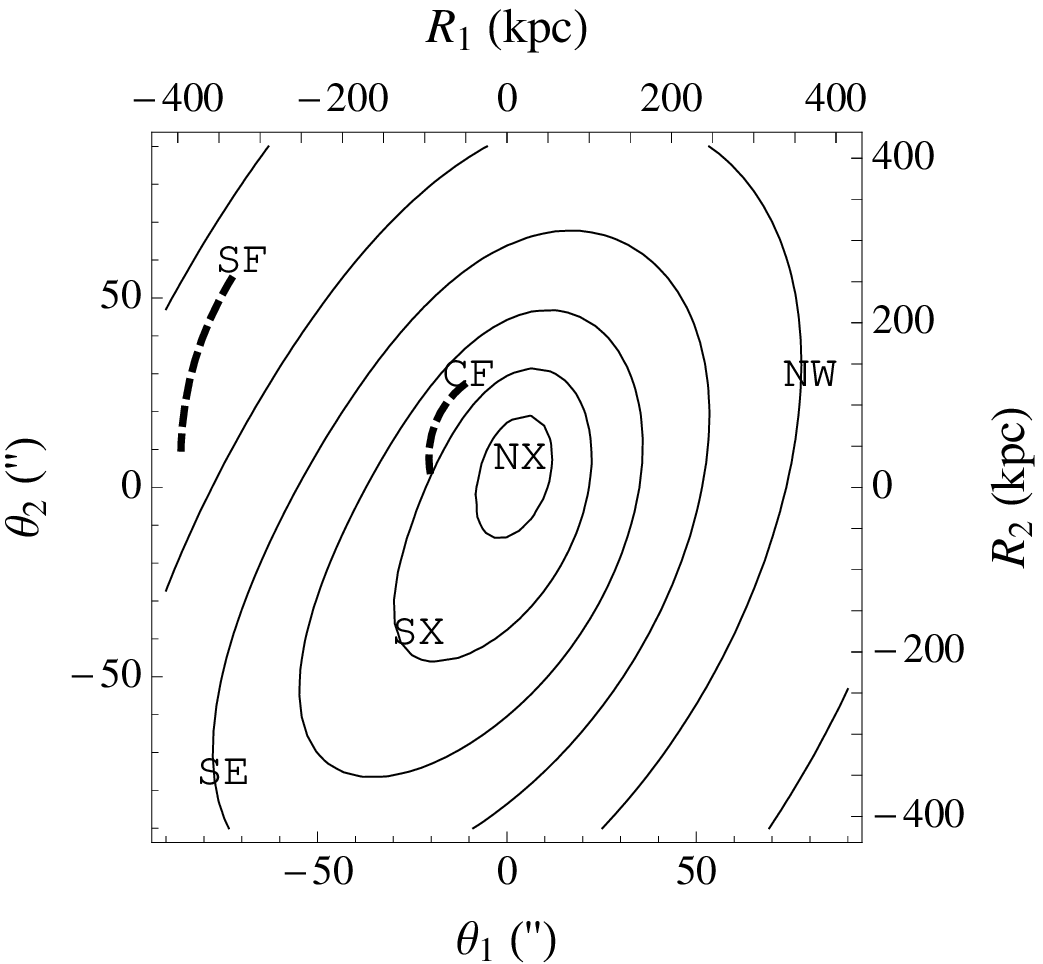}  &  
\includegraphics[width=7cm]{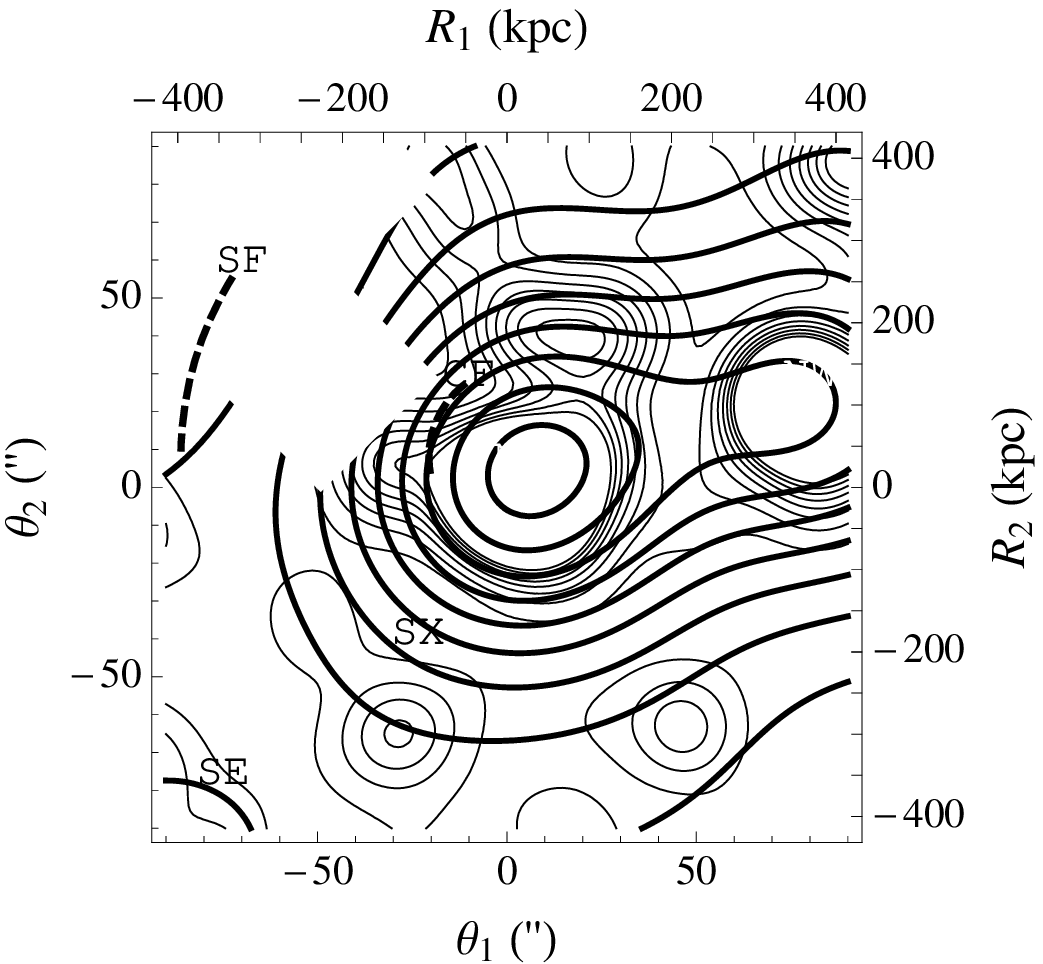} \\
\end{array}$
\end{center}
\caption{Surface matter density distribution in the core region of AC~114, in units of the projected critical density for a source redshit at $z_\mathrm{s}=3.347$, $\Sigma_{cr} = 2154.2 M_\odot\mathrm{pc}^{-2}$ ($h=0.7$). Contours represent linearly spaced values of the convergence $\kappa$. NW and SE denote the positions of two galaxy clumps; NX and SX mark the location of the X-ray surface brightness peak for the main clump and the tail, respectively. The cold front CF and the shock front SF are plotted as dashed lines. {\it Top-left}: Contour plot of the total matter density as inferred from lensing. $\kappa$-contours are plotted from 0.1 to 0.8 with a step $\Delta\kappa =0.1$. {\it Top-right}: contours of the cluster-sized dark matter halo density as derived from the lensing analysis. $\kappa$-contour values go from 0.1 to 0.7 with a step $\Delta\kappa =0.1$. {\it Bottom-left}: projected gas mass density as derived from X-ray observations. The convergence contours run from 0.07 to 0.12 with a step $\Delta\kappa =0.01$.  {\it Bottom-right}: map of the projected mass density in stars as derived from galaxy luminosities. The thick (thin) contours have been obtained by smoothing the stellar mass density with a Gaussian kernel with dispersion of $30\arcsec$ ($10\arcsec$). $\kappa$-values run from $0.2\times 10^{-2}$ to $1.8\times 10^{-2}$ with a step of $0.2\times 10^{-2}$ for the thick contours and from $2.5\times 10^{-3}$ to $2.0\times 10^{-2}$ with a step of $2\times 10^{-3}$ for the thin contours.}
\label{fig_den_mass}
\end{figure*}

The galaxy catalogue we made use of is taken from \citet[see their table~4]{cou+al98}, who morphologically classified galaxies recorded on images taken with WFPC2 at HST down to $R_{702}=23.00$. The galaxy distribution in the inner regions is quite irregular, with substructures detected with different methods, see App.~\ref{app_subs}.

Here we want to perform some analysis on the galaxy density distribution. We consider either the luminosity or the number density. Our method is as follows. We first smooth the spiky density distribution convolving with a Gaussian kernel whose fixed width is based on the mean galaxy distance in the region of interest. The features of the surface distribution are then obtained by considering a sample of maps generated resampling data by the original distribution. This takes care of the finite size error. For each map, we perform a parametric fit with Poisson weights to an elliptical density distribution. Parameter central values and confidence intervals are finally obtained considering median and quantile ranges of the final population of the sets of best fit parameters. Throughout the paper, we consider a coordinate system in the plane of the sky, $\{ \theta_1, \theta_2\}$, centred on the BCG galaxy and aligned with the equatorial system with positive numbers being to the west and north of the central galaxy. As surface density model, we consider a projected King-like $\beta$ distribution,
\beq
\label{beta1}
\Sigma = \Sigma_0 \left[ 1+(\theta_\mathrm{ell}/\theta_\mathrm{P})^2\right]^{(1-3\beta)/2} + \Sigma_\mathrm{B},
\eeq
where $\theta_\mathrm{ell}$ is a projected elliptical radius, which measures the major axis length of concentric ellipses in the plane of the sky centred in $\{\theta_{1,0},\theta_{2,0}\}$, with ellipticity $\epsilon$, defined as 1 minus the ellipse axial ratio, and orientation angle $\theta_\epsilon$ (measured North over East); $\theta_\mathrm{P}$ is the projected core radius, $\beta$ parameterizes the slope and $\Sigma_\mathrm{B}$ is the constant background. Each parameter in Eq.~(\ref{beta1}) is free to vary in the fitting procedure; flat priors are used. We consider two circular regions in the sky with external radius $\theta_\mathrm{max}=60\arcsec$ and $\theta_\mathrm{max}=120\arcsec$, containing $104$ and $329$ galaxies, respectively. For the dispersion of the Gaussian kernel we used $10\arcsec$ and $12\arcsec$, respectively. Results are listed in Table~\ref{tab_gal_den}. The central BCG galaxy is slightly shifted from the luminosity centroid, located northwest, but the statistical significance of such a displacement is low. Most notably we retrieve a significant southeast-northwest elongation. Differences between the luminosity and number density maps are due to the abundance of late-type galaxies outside the core region. Ellipticity and orientation of the luminosity distribution within $2\arcmin$ strictly follow the ellipticity parameters of the cD galaxy, which we estimated using the {\it ellipse} task in the IRAF package to be $\{\epsilon, \theta_\epsilon\} =\{0.42 \pm 0.05, (-32.8\pm0.5)\deg\}$.

\subsection{Stellar mass}
\label{sec_stel}

Baryonic contribution in stars and stellar remnants can be estimated by converting galaxy luminosities into stellar masses. We convert $R_{702}$ into infrared $K$ luminosity, which is less sensitive to ongoing star formation and is a more reliable tracer of stellar mass distribution. We mostly followed \citet{smi+al02,smi+al05}. As a first step, we corrected $R_{702}$ photometry reported in the SEXTRACTOR catalogue of \citet{cou+al98} for background overestimate as discussed in \citet{smi+al05} and then we converted $R_{702}$ photometry to Cousin $R$, using suitable corrections per morphological type \citep{smi+al02}. Then we obtained $K$ magnitudes subtracting the typical $(R-K)$ colours corrected for reddening for cluster ellipticals and spirals \citep{smi+al02}. Finally, we converted to rest-frame luminosities adopting $M_{K\odot}=3.28$ \citep{bi+me98}, Galaxy extinction of $A_K=0.023$ \citep{sch+al98} and using $K$-corrections from \citet{man+al01}. 

To convert stellar luminosity in stellar mass we followed \citet{lin+al03}. For ellipticals, we took the estimates for the central mass-to-light ratio as a function of galaxy luminosity from \citet{ger+al01}; for spiral galaxies, we used the values in \citet{be+jo01}. Estimating the mass-to-light ratios is the major source of uncertainty. Different modellings of stellar populations predict stellar mass-to-light ratios as different as $0.7~M_\odot/L_\odot$ and $1.3~M_\odot/L_\odot$ \citep{col+al01}. Additional errors are due to either interlopers included in the catalogue or missed member galaxies. Furthermore, we did not consider stars contributing to the intracluster light, whose total fraction in AC~114 is $(11\pm2)\%$ in $r$ and $(14\pm 3)\%$ in $B$ \citep{kr+be07}. It is then safe to consider an overall uncertainty $\gs 40\%$. The projected mass density in stars is plotted in Fig.~\ref{fig_den_mass}. We performed the same kind of analysis described above for the number/luminosity density. Alike to the light density, the distribution of the mass in stars is elongated from northwest to southeast. The resulting integrated mass profile in the inner core is plotted in Fig.~\ref{fig_mass_ins}. The parameters of the distribution modelled as a King profile are reported in Table~\ref{tab_gal_den}.

\section{Dynamics}
\label{sec_dyna}

\subsection{Viral mass}
\label{sec_viri}

A dynamical estimate of the total mass can be derived using the virial theorem. Assuming the cluster to be approximately spherical, non rotating and in equilibrium, the viral mass can be expressed as \citep{bi+tr87}
\beq
\label{vir1}
M_\mathrm{V}=\frac{3\pi}{2}\frac{\sigma_\mathrm{los}^2 R_\mathrm{PV}}{G} -C_\mathrm{Pr},
\eeq
where $R_\mathrm{PV}$ is the projected virial radius of the observed sample of $N$ galaxies,
\beq
\label{vir2}
R_\mathrm{PV}= \frac{N(N-1)}{\sum_{i>j}R_{ij}^{-1}},
\eeq
with $R_{ij}$ being the projected distance between galaxies $i$ and $j$. The surface term $C_\mathrm{Pr}$ accounts for the fact that the system is not entirely enclosed in the observational sample. On average, for clusters observed out to an aperture radius of $1.5~\mathrm{Mpc}/h$, the correction due to $C_\mathrm{Pr}$ is $\sim 16\%$ \citep{biv+al06}.

Combining information on galaxy position and velocity, we derived estimates of $z_\mathrm{cl}=0.3153 \pm 0.0007$ for the cluster mean redshift and $\sigma_\mathrm{los} = 1900 \pm 100~\mathrm{km~s}^{-1}$ for the cluster velocity dispersion, see App.~\ref{app_vel}. Non-members can strongly affect the mass estimate. Inclusion of interlopers that are currently infalling toward the cluster along a filament causes the overestimate of the harmonic mean radius, and, at the same time, the underestimate of the velocity dispersion. Using early-type galaxies as tracers might substantially reduce the interloper contamination in the virial mass estimate \citep{gi+me01,biv+al06}. In our approach, we accounted for this issue by estimating the interloper fraction statistically. The error on $R_\mathrm{PV}$ was estimated applying a statistical jackknife to the galaxy sample that passed the shifting gapper cut. The estimate of the cluster mass is then $M_{200} = (3.4 \pm 0.8)\times 10^{15} M_\odot/h$.

An alternative mass estimator can be based entirely on the line-of-sight velocity dispersion. As inferred from fitting to simulated clusters, the $M_{200}-\sigma$ scaling relation is  remarkably independent of cosmology. Using a cubic relation, \citet{biv+al06} obtained
\beq
\label{vir3}
M_{200}=(1.50\pm0.02)\left( \frac{\sqrt{3} \sigma_\mathrm{los}}{10^3~\mathrm{km~s}^{-1}}\right)^3 \times 10^{14}h^{-1} M_\odot .
\eeq
The intrinsic velocity distribution of early-type galaxies may be slightly biased relative to that of the dark matter particles \citep{biv+al06}, so that when using Eq.~(\ref{vir3}) it is safer to not distinguish among morphological types. To properly apply the $M_{200}-\sigma$ in Eq.~(\ref{vir3}), we correct our estimate of the intrinsic velocity dispersion, which was obtained within a given observational aperture, according to the prescription in \citet{biv+al06}. Eventually, we get $M_{200} = (4.8 \pm 0.8 \pm 0.06)\times 10^{15}  M_\odot/h$, where the second error is due to the theoretical uncertainty in the relation. We will follow this convention throughout the paper. We can see how the two mass estimates of $M_{200}$ are in agreement within the errors.

The concentration parameter can be estimated using scaling relations fitted to numerical simulations as well. According to the scaling $c_{200} \sim 4 [\sigma_\mathrm{los}/(700~\mathrm{km~s}^{-1})]^{-0.306}$ \citep{nav+al97,biv+al06}, with $\sigma_\mathrm{los}$ estimated within an aperture radius of $1.5~\mathrm{Mpc}/h$, we get $c_{200} \sim 2.95 \pm 0.05$.

\section{X-ray observations}
\label{sec_xray}

\begin{table*}
\centering
\begin{tabular}[c]{lr@{$\,\pm\,$}lr@{$\,\pm\,$}lr@{$\,\pm\,$}lr@{$\,\pm\,$}lr@{$\,\pm\,$}lr@{$\,\pm\,$}lr@{$\,\pm\,$}l}
        \hline
        \noalign{\smallskip}
        Component & \multicolumn{2}{c}{mass scale} & \multicolumn{2}{c}{$\theta_{1,0}$} & \multicolumn{2}{c}{$\theta_{2,0}$}& \multicolumn{2}{c}{$\epsilon$}
&\multicolumn{2}{c}{$\theta_\epsilon$} & \multicolumn{2}{c}{length scale}&\multicolumn{2}{c}{$\beta$} \\
        \noalign{\smallskip}
        & \multicolumn{2}{c}{} & \multicolumn{2}{c}{($\arcsec$)} &\multicolumn{2}{c}{($\arcsec$)} & \multicolumn{2}{c}{} &
\multicolumn{2}{c}{($\deg$) }
&\multicolumn{2}{c}{($\arcsec$)}&\multicolumn{2}{c}{}\\
        \noalign{\smallskip}
        \hline
         Main Clump & $0.100$&$0.008$& $3.7$&$1.0$ & $8.7$&$1.0$ & $0.39$&$0.05$ & $-13$&$4$ & $3.6$&$1.0$ & $0.389$&$0.007$\\
        \noalign{\smallskip}
        Southern Tail & $0.033$&$0.006$& $-23.4$&$1.0$ & $-37.6$&$1.0$ & $0.50$&$0.02$ & $-37$&$2$ & $170$&$40$ & $1.8$&$1.0$\\
        \noalign{\smallskip}
        \hline
\end{tabular}
\caption{Properties of the projected gas distribution as inferred from the X-ray analysis. Each component has been modelled as an isothermal $\beta$-ellipsoid.}
\label{tab_icm}
\end{table*}

AC~114 has a strongly irregular X-ray morphology \citep{def+al04}, see Fig.~\ref{fig_ac114_iso}. The cluster does not show a single X-ray peak. Noticeable emission is associated with the cluster cD galaxy but the centroid of the overall X-ray emission is located about $10\arcsec$ northwest from the cD galaxy. Two main components stand out: the cluster, roughly centered on the optical position, and a diffuse filament which spreads out towards south-east for approximately $1.5\arcmin$ ($\sim 0.3~\mathrm{Mpc}/h$), connecting the cluster core with the location of the SE clump, see App.~\ref{app_subs}.

Further signs of dynamical activity are observed north-east close to the cluster centre, see Fig.~\ref{fig_den_mass}: a cold front at $\sim 20\arcsec$ ($\sim 70~\mathrm{kpc}/h$) from the core center and a likely shock front at $\sim 90 \arcsec$ ($\sim 0.3~\mathrm{Mpc}/h$). 
%A second hard region is observed on the western side of the cluster core opposite to the shock front, but drastic jumps in the surface brightness and temperature are not observed there. The location of the cold front might be compatible with a sloshing core \citep{ti+he05,ma+vi07}, but the single front, the hard ratio regions and the filament are more consistent with an ongoing major merger. 

The tail and the fronts might be independent phenomena. Diffuse X-ray emission is detected near the SE clump, whereas no X-ray emission is associated to the NW clump. The NE substructure detected with the $\delta_\mathrm{DS}$-test, see App.~\ref{app_subs}, was not targeted by X-ray observations. One possible scenario is that the SE clump, in its motion from the northwest through the cluster, has been ram-pressure stripped of most of its intra-group gas, now still visible as the soft southern tail. The interaction with the cluster might have also caused the asymmetrical stretch of the cluster emission detected toward south-east. The NW clump might have been stripped as well.

%The fronts suggest the motion of a sub-structure toward north-east which compresses and heats intra-cluster gas ahead of its path. Unfortunately, such a bullet is expected to be outside the region targeted by our strong lensing analysis, see Sec.~\ref{sec_stro}, so that we can not confirm this scenario independently. Shock fronts associated with moving sub-structures are short-living phenomena, and are therefore signs of recent merging processes.  From the Mach number ($=1.3\pm0.1$) and the post-shock gas temperature ($T=7.0^{+1.5}_{-1.0}$) measured in \citet{def+al04}, we derive a shock velocity of $\sim 1800~\mathrm{km~s}^{-1}$.  Assuming that the bullet subcluster velocity is close to the shock velocity and that the distance between the centres of the merging sub-clusters is of the order of the shock clustercentric distance, the two subclusters would have passed through each other just 0.15 Gyr ago. 

The cluster bolometric luminosity is $L_X =(6.7^{+0.4}_{-0.2})\times 10^{44}~\mathrm{erg~s}^{-1}/h^2$ \citep{def+al04}. Based on scaling relations between luminosity and velocity dispersion \citep{ryk+al08}, we would expect $L_X =(5 \pm 1 \pm 1)\times 10^{45}~\mathrm{erg~s}^{-1}/h^2$, larger than the observed value. This might suggest that on one hand the gas has still to settle down in the cluster potential well and, on the other hand, the clumpy structure of AC~114 might bring about an overestimate of the velocity dispersion.

\subsection{Projected mass density}

The gas mass can be estimated from the X-ray emission. The surface brightness was modelled in \citet{def+al04} as a sum of two elliptical isothermal $\beta$-profiles, a main core plus an extended off-centred south-east tail. In order to infer the projected mass associated to each component, we have to project the corresponding 3D ellipsoid, which were previously obtained by de-projecting the observed intensity map, see App.~\ref{app_proj}. The resulting projected density profile for each component follows a  King-like $\beta$ distribution, see Eq.~(\ref{beta1}); parameters are listed in Table~\ref{tab_icm}. In order to obtain the corresponding three-dimensional electron density, we took care of deprojecting the surface brightness maps of the main clump and the tail separately. Then, we projected back the two separate distributions to the lens plane, see App.~\ref{app_proj} for details. Since we treated the main clump and the tail separately, the only uncertainty on the projected mass due to the casting method is then a correction geometrical factor which depends on the unknown intrinsic axial ratios and orientation angles. The projected mass map is plotted in Fig.~\ref{fig_den_mass}. The integrated ICM mass is plotted in Fig.~\ref{fig_mass_ins}. 

The main sources of error for the gas mass are the projection effects and the assumption of isothermal emission. Missing knowledge of intrinsic axial ratios and orientation of the gas distribution brings about an uncertainty in the overall normalization of the projected gas density. In App.~\ref{app_proj}, we estimate such an uncertainty to be of the order of $\sim 18\%$. 

As expected from the several signs of dynamical activity, there is no evidence for a central cool core. The radial profile suggests a decline at large radii \citep[see their figure~6]{def+al04}, but due to the large errors a constant temperature is in full agreement with data. Furthermore, in the small central region targeted by the strong lensing analysis ($\ls 100$~kpc), there is no evidence for radial variations. Then the error due to deviations from isothermality, which can be estimated to be of a few percents, is much smaller than the uncertainty on the spectroscopic determination of the temperature.

In the inner core, the contribution of the tail is subdominant. Note that the central convergence for the ICM is $\kappa_0 \ls 0.13$, so that the gas mass is subcritical for lensing.

\section{Strong lensing analysis}
\label{sec_stro}

\begin{figure*}
	\centering
		\includegraphics[width=0.7\textwidth]{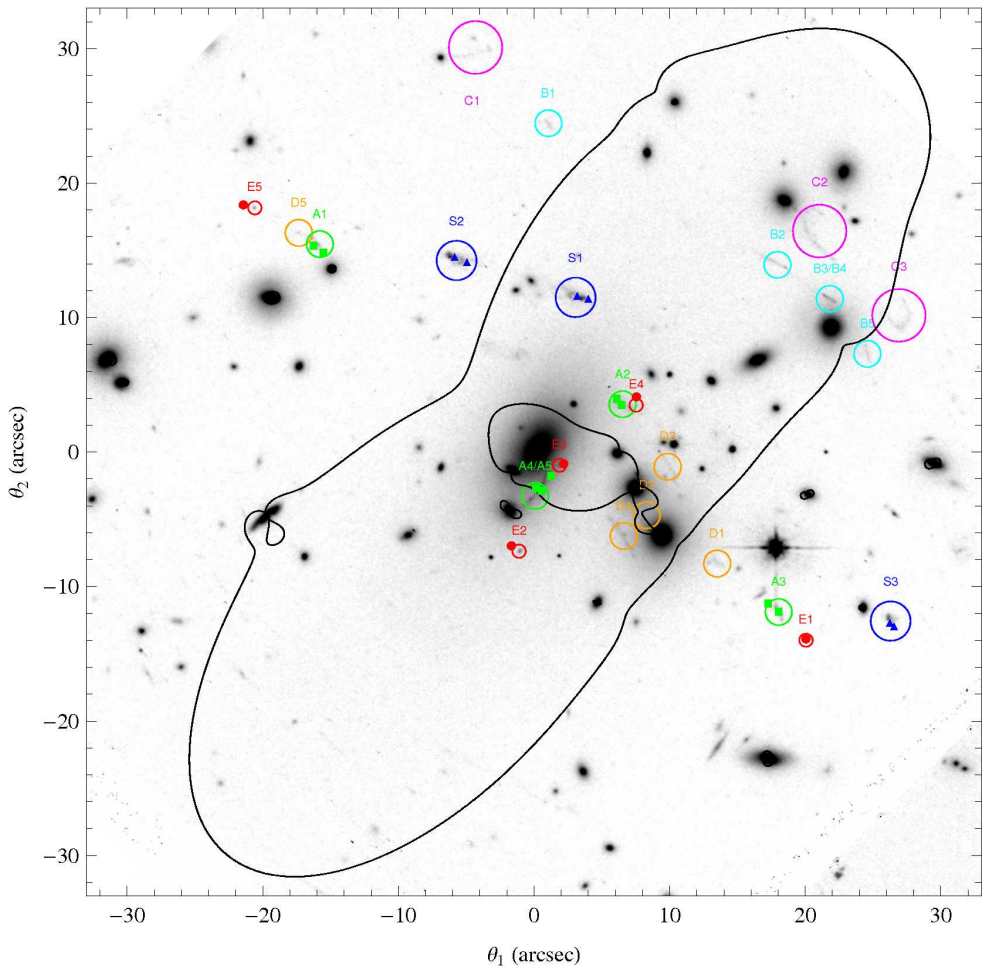}
		\caption{HST/WFPC-2 image of AC~114 with the observed multiple image systems. The coordinates $\theta_1$ and $\theta_2$, both measured in arcseconds are in direction west and north, respectively. The critical lines are represented by the black lines and are referred to the image system E ($z_\mathrm{s}=3.347$). Circles surround multiple images (A, B, C, D, E and S systems), while the filled squares, small circles and triangles mark the predicted image positions for the system A (green in the electronic version of the paper), E (red) and S (blue), respectively.}
		\label{fig_crit_img}
\end{figure*}

\subsection{Optical data}

\begin{figure}
	\centering
		\subfigure{\fbox{\includegraphics[width=0.14\textwidth]{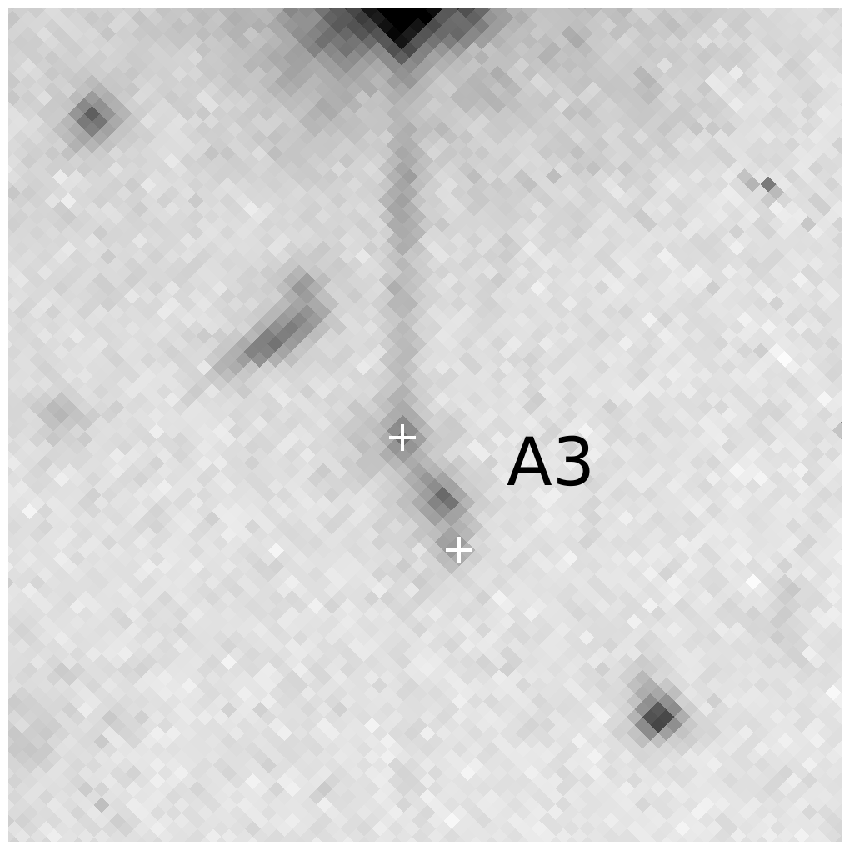}}}
		\subfigure{\fbox{\includegraphics[width=0.14\textwidth]{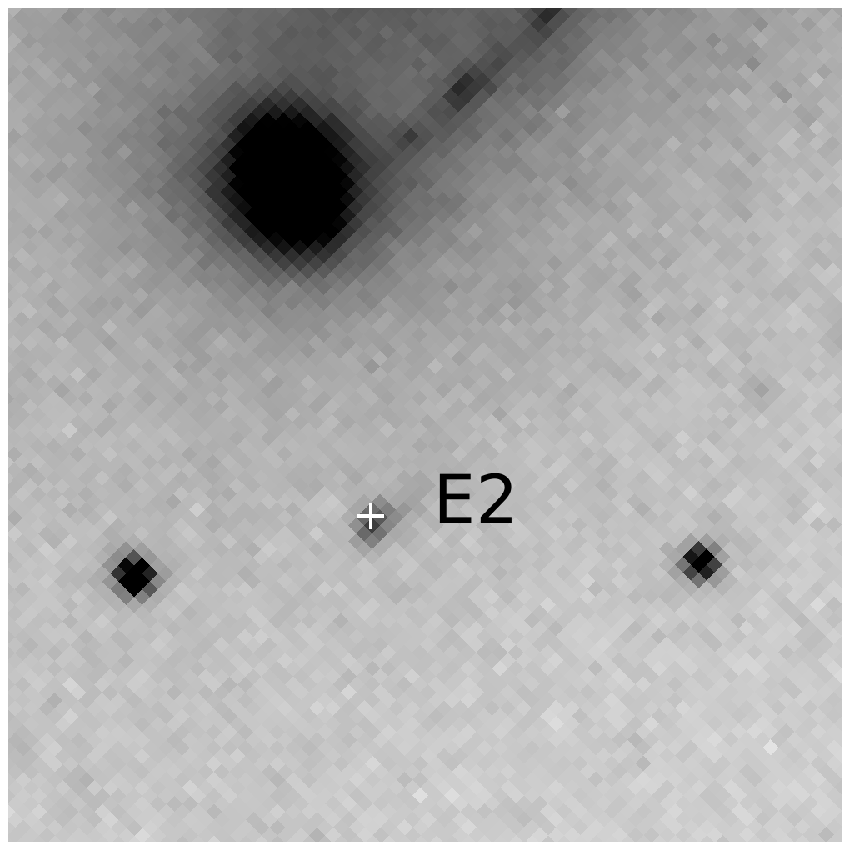}}}
		\subfigure{\fbox{\includegraphics[width=0.14\textwidth]{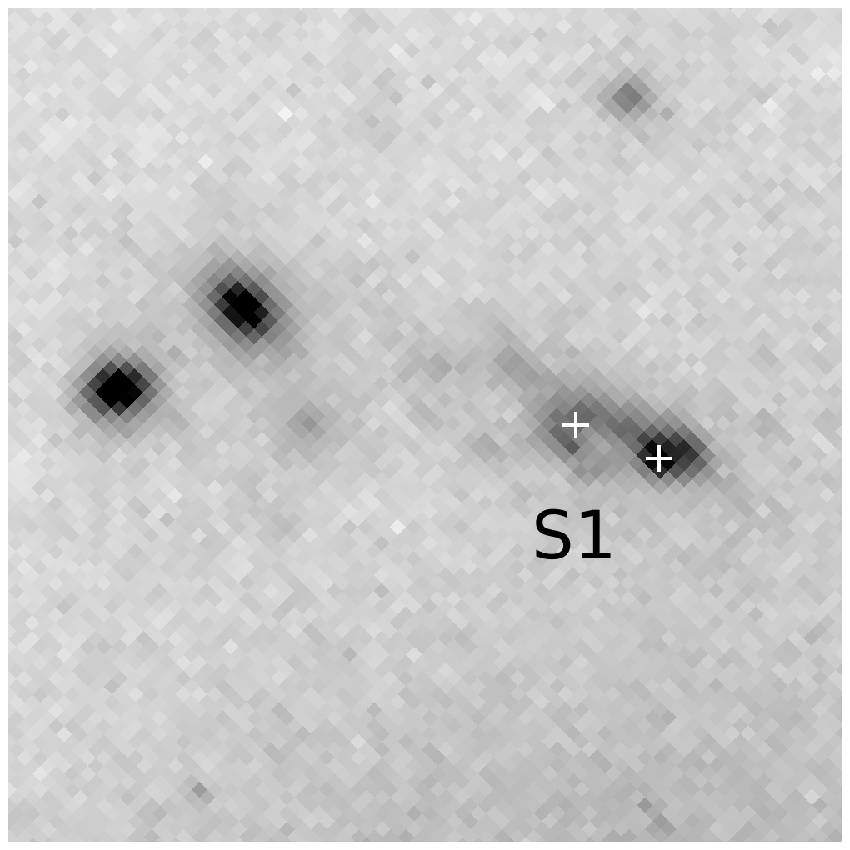}}}
	\caption{A mosaic of the zoomed-in regions ($\sim 8\arcsec \times 8\arcsec$) surroundings the images A3 (left panel), E2 (middle panel) and S1 (right panel). A3 has an elongated shape wherein distinct points can be recognized. For E2, the intrinsic morphology of the source can not be distinguished, whereas system S images are hook-shaped, as shown here for S1. The white crosses are the coordinates of the sampled points used into the strong lensing analysis.}
	\label{fig_img_AES}
\end{figure}

Many multiple image systems have been detected in the core of AC~114, see Fig. \ref{fig_crit_img}. The first ones were discovered in a survey for bright gravitational lensing arcs by \citet{sef91}. Two images of the prominent three-image system S were first identified by \citet{sce95}, whereas the third image S3 and the systems A, B, C and D were discovered by \citet{nat+al98}. The last image system E has been located by \citet{cam+al01}, who also measured source redshifts through spectroscopic observations.

For our strong lensing model, we exploited only the image systems with confirmed spectroscopic redshift, i.e. A, E and S. The other systems have not been considered, as they are strongly perturbed by some cluster galaxies or lack precise redshift measurements. The image system S, at redshift $z_\mathrm{s}=1.867$, is composed of three hook-shaped images, see Fig.~\ref{fig_img_AES}. In order to take into account the parity and the orientation of the images and to exploit the information carried by the shape, each S-image was sampled by two points. We considered an uncertainty of $0.4\arcsec$, which will be the default error for each positional data. The image system E is composed of five nearly point-like images at redshift $z_\mathrm{s}=3.347$, see Fig.~\ref{fig_img_AES}. The multiple image system A is composed of five images of a single source at redshift $z_\mathrm{s}=1.691$. The images A1, A2 and A3 are only weakly stretched by the lens and it can be seen morphologically that they are images of the same source. We distinguished two conjugate knots in each image, see Fig.~\ref{fig_img_AES}. On the other hand, A4 and A5 are strongly stretched because they are merging into a single arc across the radial critical curve near the BCG. As the knots in these two central images can not be distinguished, they have been furnished with a larger uncertainty ($1.6\arcsec$). 

The adopted positional uncertainties are larger than the HST astrometric resolution. Clusters are complex systems and simple models can not account for all the mass complexities. A coarser positional error allows to perform the lensing analysis without adding too much parameters and, at the same time, avoiding that the region in parameter space explored is overly confined \citep{san+al08}. This approach can be effective when dealing with galaxy clumps as those revealed by the AC~114 luminosity map, which are usually associated with meso-structures \citep{sah+al07}.

\subsection{Mass components}
\label{subsec-models}

We performed a strong lensing analysis which exploits optical observations, see Sec.~\ref{sec_gala}, as well as measurements in the X-ray band, see Sec.~\ref{sec_xray}. This multi-wavelength approach allowed us to model the three main components: the cluster-sized dark matter halo, the cluster-sized ICM and the observed galaxies. Each component was described with a separate parametric mass model.

The projected surface mass density $\Sigma$ of these density profiles is expressed in terms of the convergence $\kappa$, i.e. in units of the critical surface mass density for lensing, $\Sigma_\mathrm{cr}=(c^2\,D_\mathrm{s})/(4\pi G\,D_\mathrm{d}\,D_\mathrm{ds})$, where $D_\mathrm{s}$, $D_\mathrm{d}$ and $D_\mathrm{ds}$ are the source, the lens and the lens-source angular diameter distances, respectively. We considered mass distributions with elliptical symmetry, so that the convergence can be written in terms of the elliptical radius $\theta_\mathrm{ell}$. 

To model the cluster-sized DM component, we considered parametric mass models with either isothermal or Navarro-Frenk-White (NFW) density profiles. DM halos are successfully described as NFW profiles \citep{nfw96,nav+al97}, whose 3D distribution follows
\begin{equation}
	\rho_\mathrm{NFW}=\frac{\rho_\mathrm{s}}{(r/r_\mathrm{s})(1+r/r_\mathrm{s})^2},
\end{equation}
where $\rho_\mathrm{s}$ is the characteristic density and $r_\mathrm{s}$ is the characteristic length scale. The convergence for this mass profile with elliptical symmetry is obtained by replacing the polar with the elliptical radius in the resulting projected surface mass density. We will describe the projected NFW density in terms of the strength of the lens $\kappa_\mathrm{NFW}$, see Eq.~(\ref{nfw1}), and of the projected length scale $r_\mathrm{sP}$, see App.~\ref{app_proj}, i.e. the two parameters directly inferred by fitting projected lensing maps.

An alternative description for a DM component is in terms of isothermal mass density. The non-singular isothermal profiles are parametrized by a softened power-law ellipsoid (NIE), and represent a special case of $\beta$-models with $\beta=2/3$, see Eq.~(\ref{beta1}). The mass scale parameter is usually written as $b=2 \kappa_0 r_\mathrm{cP}$ \citep{kee01a}, where $\kappa_0$ is the central convergence and $r_\mathrm{cP}$ is the projected core radius.

The two gas components, i.e. the main X-ray clump and the soft tail, can be modelled as $\beta$-profiles, see Eq.~(\ref{beta1}). Unlike the DM component, which was modelled as isothermal, the slope for each gas component is fixed by the X-ray observations, see Table~\ref{tab_icm}. We note that the mass distribution of the main X-ray emitting clump is quite flat, so that the impact on lensing features is limited. We considered a normal prior on the mass normalization $\kappa_0$ of the main gas component with mean and dispersion as inspired from the X-ray analysis, see Table~\ref{tab_icm}, and sharp priors on the remaining parameters describing the ICM distribution.

For an accurate lens modeling, the mass distribution of the galaxies has to be considered too. Galaxies are small compared to the whole cluster, but they have high local mass densities and can strongly perturb the cluster potential in their neighborhood. Therefore, galaxies which affect the considered image systems are taken into account. The selection has been limited to the region of the cluster where the multiple image systems are located. We selected galaxies brighter than $R_{702}=21.2$ within a radius of $44\arcsec$ from the BCG; 25 galaxies passed the cut.

Galaxy-sized halos can be modeled by pseudo-Jaffe mass profiles, which are obtained subtracting a NIE with core radius $r_\mathrm{t}$ (called truncation radius) from another NIE with core radius $r_\mathrm{c}$, where $r_\mathrm{c}<r_\mathrm{t}$. Apart from the BCG, see Sec.~\ref{sec_gala}, we considered spherical galaxies. Each pseudo-Jaffe model is characterised by a velocity dispersion $\sigma_\mathrm{DM}$, a core radius $r_\mathrm{c}$ and a truncation radius $r_\mathrm{t}$. To minimize the number of parameters, a set of scaling laws has been adopted \citep{bbs96}: $\sigma_\mathrm{DM}=\sigma_\mathrm{DM}^*\left( L/L^*\right)^{1/4}$ and $r_\mathrm{t}=r_\mathrm{t}^*\left(L/L^*\right)^{1/2}$. The core radius $r_\mathrm{c}$ is scaled in the same way as $r_\mathrm{t}$. The dispersion $\sigma_\mathrm{DM}$ is related to the total mass through $M=(9/2 G)\sigma_\mathrm{DM}^2 r_\mathrm{t}$ \citep{nat+al98}. As characteristic luminosity $L^*$ we considered an hypothetical galaxy with $R_{702}=19.5$. We considered a flat prior on $\sigma_\mathrm{DM}^*$, which was left free to vary between $0$ and $450~\mathrm{km/s}$, whereas $r_\mathrm{c}^*$ and $r_\mathrm{t}^*$ were fixed to $0.15~\mathrm{kpc}$ and $45\,\mathrm{kpc}$, respectively \citep{nat+al09}. 

BCGs are a distinct galaxy population from $L^*$ cluster ellipticals and should be modelled by their own. Scaling the BCGs as average cluster members would be ineffective when studying mass-to-light ratios of typical early-type galaxies \citep{nat+al98}. However, as far as ellipticity, orientation, and centroid  (the main features we are going to compare among the different mass components) of the cluster-sized DM halo are concerned, a different lensing modelling of the BCG would has no significant impact. 

In general, modelling each perturbing galaxy by its own would help to obtain a better fit to the data \citep{lim+al08}. On the other hand, the cluster-sized DM halo cannot be effectively distinguished by the BCG one through pure lensing analyses.\footnote{Dynamical analyses of the inner velocity profile are required to disentangle the cluster-sized from the BCG contribution \citep{san+al08}.} As far as a regular cluster is concerned, one could assume that the two halos are centered at the same position and then model only the stellar content of the BCG in the lensing model. In such a way, the cluster-sized DM halo would account also for the BCG dark halo. 

We wanted to explore the mass components in AC~114 without forcing the DM distribution to follow either the gas or the galaxy density and we left the DM centroid free. Since a reasonable physical model requires DM associated to the BCG, we had to account also for the BCG halo. However, in the present paper, we were mainly concerned with the cluster-sized DM component so that we preferred to keep the number of free parameters linked to galactic halos as small as possible. We then considered three different modellings. As a first case, the BCG was modelled on its own as a pseudo-Jaffe profile with ellipticity fixed by his luminosity distribution and velocity dispersion modelled after imposing a flat prior $0 < \sigma_\mathrm{DM} \le 500\mathrm{km/s}$. Alternatively, we forced the BCG total mass to follow the same scaling relations as the other galaxies. With such a scaling, the NFW cluster-sized profile makes up for most of the DM associated with the BCG halo. In both cases, the core and the truncation radius were scaled according to the characteristic values. This has a negligible effect due to the degeneracy between the scale-length and the velocity dispersion. As a final case we let the total BCG mass distribution to be embedded in the cluster-sized dark matter halo. Note that this worked only for the cusped NFW halo. 

We remark that our analysis is not meant to investigate if the BCG can either be included as part of the main cluster or as a separate potential. In fact, the above tested models differ in the values of the central velocity dispersion profile and should be distinguished by exploiting dynamical analyses in the very inner regions. We just considered such very different cases of BCG modelling to show that the impact of gas in lens modelling is nearly independent of the galaxies. However, we stress that gas and star mass distribution, discussed in Sec.~\ref{sec_resu}, were inferred with tools independent of the lensing analysis.

\section{Inferred mass distribution}
\label{sec_infe}

\begin{table*}
\centering
\begin{tabular}[c]{lr@{$\,\pm\,$}lr@{$\,\pm\,$}lr@{$\,\pm\,$}lr@{$\,\pm\,$}lr@{$\,\pm\,$}lr@{$\,\pm\,$}lr@{$\,\pm\,$}ll}
        \hline
        \noalign{\smallskip}
        Component & \multicolumn{2}{c}{mass scale} & \multicolumn{2}{c}{$\theta_{1,0}$} & \multicolumn{2}{c}{$\theta_{2,0}$}& \multicolumn{2}{c}{$\epsilon$}&\multicolumn{2}{c}{$\theta_\epsilon$} & \multicolumn{2}{c}{length scale}&\multicolumn{2}{c}{} & $\log E$\\
        \noalign{\smallskip}
        & \multicolumn{2}{c}{} & \multicolumn{2}{c}{($\arcsec$)} &\multicolumn{2}{c}{($\arcsec$)} & \multicolumn{2}{c}{} &
\multicolumn{2}{c}{($\deg$) }&\multicolumn{2}{c}{($\arcsec$)}&\multicolumn{2}{c}{}&\\
        \noalign{\smallskip}
        \hline
        \noalign{\smallskip}
        \bf{All} & \multicolumn{2}{c}{$\kappa_\mathrm{NFW}$} & \multicolumn{8}{c}{}
&\multicolumn{2}{c}{$r_\mathrm{sP}$}&\multicolumn{2}{c}{}& \\
        \noalign{\smallskip}
        NFW & $0.213$&$0.017$& $1.50$&$0.17$ & $-0.5$&$0.2$ & $0.50$&$0.03$ &
$-38.3$&$0.4$ & $210$&$30$& \multicolumn{2}{c}{-}&-12.4\\
\noalign{\smallskip}
\hline

      \bf{Halo with ICM} & \multicolumn{2}{c}{$\kappa_\mathrm{NFW}$} & \multicolumn{8}{c}{}
&\multicolumn{2}{c}{$r_\mathrm{sP}$}&\multicolumn{2}{c}{}&\\
        \noalign{\smallskip}
        NFW & $0.195$&$0.017$& $1.0$&$0.3$ & $-0.6$&$0.3$ & $0.54$&$0.03$ & $-37.2$&$0.6$
& $240$&$40$ & \multicolumn{2}{c}{-}&-9.2\\
        \noalign{\smallskip}
        \bf{All galaxies} &
\multicolumn{2}{c}{$\sigma_\mathrm{DM}~(\mathrm{km}~\mathrm{s}^{-1})$} &
\multicolumn{8}{c}{}&\multicolumn{2}{c}{$r_\mathrm{cP}~(\mathrm{kpc})$}&\multicolumn{2}{c}{$r_\mathrm{tP}~(\mathrm{kpc})$}&\\
        \noalign{\smallskip}
        pJaffe-$L^*$ & $85$&$19$ & \multicolumn{2}{c}{-} & \multicolumn{2}{c}{-} &
\multicolumn{2}{c}{-} & \multicolumn{2}{c}{-} & \multicolumn{2}{c}{$[0.15]$} &
\multicolumn{2}{c}{$[45]$}&\\

\noalign{\smallskip}
\hline

\bf{Halo with ICM and BCG} & \multicolumn{2}{c}{$\kappa_\mathrm{NFW}$} & \multicolumn{8}{c}{}
&\multicolumn{2}{c}{$r_\mathrm{sP}$}&\multicolumn{2}{c}{}&\\
        \noalign{\smallskip}
        NFW & $0.205$&$0.016$& $1.1$&$0.3$ & $-0.3$&$0.3$ & $0.52$&$0.03$ & $-37.1$&$0.7$
& $220$&$30$ & \multicolumn{2}{c}{-}&-7.6\\
        \noalign{\smallskip}
        \bf{Other galaxies} &
\multicolumn{2}{c}{$\sigma_\mathrm{DM}~(\mathrm{km}~\mathrm{s}^{-1})$} &
\multicolumn{8}{c}{}&\multicolumn{2}{c}{$r_\mathrm{cP}~(\mathrm{kpc})$}&\multicolumn{2}{c}{$r_\mathrm{tP}~(\mathrm{kpc})$}&\\
        \noalign{\smallskip}
         pJaffe-$L^*$ & $90$&$20$ & \multicolumn{2}{c}{-} & \multicolumn{2}{c}{-} &
\multicolumn{2}{c}{-} & \multicolumn{2}{c}{-} & \multicolumn{2}{c}{$[0.15]$} &
\multicolumn{2}{c}{$[45]$}&\\

\noalign{\smallskip}
\hline

\bf{Halo with all galaxies} & \multicolumn{2}{c}{$\kappa_\mathrm{NFW}$} & \multicolumn{8}{c}{}
&\multicolumn{2}{c}{$r_\mathrm{sP}$}&\multicolumn{2}{c}{}&\\
        \noalign{\smallskip}
        NFW & $0.206$&$0.016$& $1.47$&$0.17$ & $-0.48$&$0.19$ & $0.51$&$0.03$ & $-40.6$&$0.6$
& $170$&$20$ & \multicolumn{2}{c}{-}&-4.0\\
        \noalign{\smallskip}
     \bf{ICM}& \multicolumn{2}{c}{$\kappa_0$} & \multicolumn{8}{c}{}
&\multicolumn{2}{c}{$r_\mathrm{cP}$}&\multicolumn{2}{c}{$\beta$}\\
        \noalign{\smallskip}
        Main Clump & $0.094$&$0.011$&  \multicolumn{2}{c}{[3.7]} & \multicolumn{2}{c}{[8.7]} & \multicolumn{2}{c}{[0.39]} & \multicolumn{2}{c}{[-13]} & \multicolumn{2}{c}{[13.6]} & \multicolumn{2}{c}{[0.389]}& \\
        \noalign{\smallskip}
        Southern Tail & \multicolumn{2}{c}{[0.033]} & \multicolumn{2}{c}{[-23.4]} & \multicolumn{2}{c}{[-37.6]} & \multicolumn{2}{c}{[0.50]} & \multicolumn{2}{c}{[-37]} & \multicolumn{2}{c}{[170]} & \multicolumn{2}{c}{[1.8]}&\\    

\noalign{\smallskip}
\hline

\bf{Halo} & \multicolumn{2}{c}{$\kappa_\mathrm{NFW}$} & \multicolumn{8}{c}{}
&\multicolumn{2}{c}{$r_\mathrm{sP}$}&\multicolumn{2}{c}{}&\\
        \noalign{\smallskip}
        NFW & $0.190$&$0.019$& $1.2$&$0.5$ & $-0.4$&$0.3$ & $0.53$&$0.03$ & $-40$&$1$
& $190$&$30$ & \multicolumn{2}{c}{-}&-1.9\\
          \noalign{\smallskip}
        \bf{All galaxies} &
\multicolumn{2}{c}{$\sigma_\mathrm{DM}~(\mathrm{km}~\mathrm{s}^{-1})$} &
\multicolumn{8}{c}{}&\multicolumn{2}{c}{$r_\mathrm{cP}~(\mathrm{kpc})$}&\multicolumn{2}{c}{$r_\mathrm{tP}~(\mathrm{kpc})$}&\\
        \noalign{\smallskip}
        pJaffe-$L^*$  & $70$&$30$ & \multicolumn{2}{c}{-} & \multicolumn{2}{c}{-} &
\multicolumn{2}{c}{-} & \multicolumn{2}{c}{-} & \multicolumn{2}{c}{$[0.15]$} &
\multicolumn{2}{c}{$[45]$}&\\
\noalign{\smallskip}
     \bf{ICM}& \multicolumn{2}{c}{$\kappa_0$} & \multicolumn{8}{c}{}
&\multicolumn{2}{c}{$r_\mathrm{cP}$}&\multicolumn{2}{c}{$\beta$}\\
        \noalign{\smallskip}
        Main Clump & $0.093$&$0.013$&  \multicolumn{2}{c}{[3.7]} & \multicolumn{2}{c}{[8.7]} & \multicolumn{2}{c}{[0.39]} & \multicolumn{2}{c}{[-13]} & \multicolumn{2}{c}{[13.6]} & \multicolumn{2}{c}{[0.389]}& \\
        \noalign{\smallskip}
        Southern Tail & \multicolumn{2}{c}{[0.033]} & \multicolumn{2}{c}{[-23.4]} & \multicolumn{2}{c}{[-37.6]} & \multicolumn{2}{c}{[0.50]} & \multicolumn{2}{c}{[-37]} & \multicolumn{2}{c}{[170]} & \multicolumn{2}{c}{[1.8]}&\\    
 
\noalign{\smallskip}
\hline

\bf{Halo with BCG} & \multicolumn{2}{c}{$\kappa_\mathrm{NFW}$} & \multicolumn{8}{c}{}
&\multicolumn{2}{c}{$r_\mathrm{sP}$}&\multicolumn{2}{c}{}&\\
        \noalign{\smallskip}
        NFW & $0.198$&$0.017$& $1.2$&$0.3$ & $-0.3$&$0.3$ & $0.53$&$0.03$ & $-39.4$&$0.9$
& $180$&$30$ & \multicolumn{2}{c}{-}&0.0\\
          \noalign{\smallskip}
        \bf{Other galaxies} &
\multicolumn{2}{c}{$\sigma_\mathrm{DM}~(\mathrm{km}~\mathrm{s}^{-1})$} &
\multicolumn{8}{c}{}&\multicolumn{2}{c}{$r_\mathrm{cP}~(\mathrm{kpc})$}&\multicolumn{2}{c}{$r_\mathrm{tP}~(\mathrm{kpc})$}&\\
        \noalign{\smallskip}
        pJaffe & $80$&$20$ & \multicolumn{2}{c}{-} & \multicolumn{2}{c}{-} &
\multicolumn{2}{c}{-} & \multicolumn{2}{c}{-} & \multicolumn{2}{c}{$[0.15]$} &
\multicolumn{2}{c}{$[45]$}&\\
\noalign{\smallskip}
     \bf{ICM}& \multicolumn{2}{c}{$\kappa_0$} & \multicolumn{8}{c}{}
&\multicolumn{2}{c}{$r_\mathrm{cP}$}&\multicolumn{2}{c}{$\beta$}\\
        \noalign{\smallskip}
        Main Clump & $0.092$&$0.013$&  \multicolumn{2}{c}{[3.7]} & \multicolumn{2}{c}{[8.7]} & \multicolumn{2}{c}{[0.39]} & \multicolumn{2}{c}{[-13]} & \multicolumn{2}{c}{[13.6]} & \multicolumn{2}{c}{[1.833]}& \\
        \noalign{\smallskip}
        Southern Tail & \multicolumn{2}{c}{[0.033]} & \multicolumn{2}{c}{[-23.4]} & \multicolumn{2}{c}{[-37.6]} & \multicolumn{2}{c}{[0.50]} & \multicolumn{2}{c}{[-37]} & \multicolumn{2}{c}{[170]} & \multicolumn{2}{c}{[-2.4]}&\\    
\noalign{\smallskip}
\hline
\end{tabular}
\caption{Model parameters inferred from the lensing analysis. For each model, the first line specifies the mass components accounted by the cluster-sized halo, parameterized as a NFW distribution.``All" refers to a mass model (DM+baryons) with a single NFW component. The other models may have additional components either for the gas or the galactic halos. The BCG is either scaled as the other galaxies (``All galaxies") or embedded in the cluster-zised halo (halos ``with BCG"). The ICM is either modelled apart or embedded in the cluster-sized halo (``Halo with ICM"). The orientation angle $\theta_\epsilon$ is measured North over East. $r_\mathrm{sP}$ ($r_\mathrm{cP}$) is the projected length scale (core radius) for the NFW ($\beta$) profile. Parameters in squared brackets were kept fixed in the lensing analysis. Convergences are normalised to the E system source redshift.}
\label{tab_best_fit}
\end{table*}

\begin{table*}
\centering
\begin{tabular}[c]{lr@{$\,\pm\,$}lr@{$\,\pm\,$}lr@{$\,\pm\,$}lr@{$\,\pm\,$}lr@{$\,\pm\,$}lr@{$\,\pm\,$}lr@{$\,\pm\,$}ll}
        \hline
        \noalign{\smallskip}
        Component & \multicolumn{2}{c}{mass scale} & \multicolumn{2}{c}{$\theta_{1,0}$} & \multicolumn{2}{c}{$\theta_{2,0}$}& \multicolumn{2}{c}{$\epsilon$}&\multicolumn{2}{c}{$\theta_\epsilon$} & \multicolumn{2}{c}{length scale}&\multicolumn{2}{c}{} & $\log E$\\
        \noalign{\smallskip}
        & \multicolumn{2}{c}{} & \multicolumn{2}{c}{($\arcsec$)} &\multicolumn{2}{c}{($\arcsec$)} & \multicolumn{2}{c}{} &
\multicolumn{2}{c}{($\deg$) }&\multicolumn{2}{c}{($\arcsec$)}&\multicolumn{2}{c}{}&\\
        \noalign{\smallskip}
        \hline
        \noalign{\smallskip}

      \bf{Halo with ICM} & \multicolumn{2}{c}{$b~(\arcsec)$} & \multicolumn{8}{c}{}
&\multicolumn{2}{c}{$r_\mathrm{sP}$}&\multicolumn{2}{c}{}&-13.7\\
        \noalign{\smallskip}
        NIE & $27$&$2$& $0.2$&$0.7$ & $-0.4$&$0.7$ & $0.59$&$0.04$ & $-35.2$&$0.6$
& $11.9$&$1.4$ & \multicolumn{2}{c}{-}&\\
        \noalign{\smallskip}
        \bf{Galaxies} &
\multicolumn{2}{c}{$\sigma_\mathrm{DM}~(\mathrm{km}~\mathrm{s}^{-1})$} &
\multicolumn{8}{c}{}&\multicolumn{2}{c}{$r_\mathrm{cP}~(\mathrm{kpc})$}&\multicolumn{2}{c}{$r_\mathrm{tP}~(\mathrm{kpc})$}&\\
        \noalign{\smallskip}
        pJaffe-$L^*$ & $147$&$13$ & \multicolumn{2}{c}{-} & \multicolumn{2}{c}{-} &
\multicolumn{2}{c}{-} & \multicolumn{2}{c}{-} & \multicolumn{2}{c}{$[0.15]$} &
\multicolumn{2}{c}{$[45]$}&\\

\noalign{\smallskip}
\hline

      \bf{Halo with ICM} & \multicolumn{2}{c}{$b~(\arcsec)$} & \multicolumn{8}{c}{}
&\multicolumn{2}{c}{$r_\mathrm{sP}$}&\multicolumn{2}{c}{}&-6.7\\
        \noalign{\smallskip}
        NIE & $31$&$3$& $0.8$&$0.8$ & $-0.6$&$0.7$ & $0.55$&$0.04$ & $-36.4$&$0.6$
& $13.7$&$1.4$ & \multicolumn{2}{c}{-}&\\
        \noalign{\smallskip}
        \bf{Galaxies} &
\multicolumn{2}{c}{$\sigma_\mathrm{DM}~(\mathrm{km}~\mathrm{s}^{-1})$} &
\multicolumn{8}{c}{}&\multicolumn{2}{c}{$r_\mathrm{cP}~(\mathrm{kpc})$}&\multicolumn{2}{c}{$r_\mathrm{tP}~(\mathrm{kpc})$}&\\
        \noalign{\smallskip}
        pJaffe-$L^*$ & $110$&$20$ & \multicolumn{2}{c}{-} & \multicolumn{2}{c}{-} &
\multicolumn{2}{c}{-} & \multicolumn{2}{c}{-} & \multicolumn{2}{c}{$[0.15]$} &
\multicolumn{2}{c}{$[45]$}&\\
pJaffe-BCG & $239$&$18$ & \multicolumn{2}{c}{[0.0]} & \multicolumn{2}{c}{[0.0]} &
\multicolumn{2}{c}{[0.42]} & \multicolumn{2}{c}{[-32.8]} & \multicolumn{2}{c}{$[0.29]$} &
\multicolumn{2}{c}{$[87]$}&\\

\noalign{\smallskip}
\hline

     \bf{Halo} & \multicolumn{2}{c}{$b~(\arcsec)$} & \multicolumn{8}{c}{}
&\multicolumn{2}{c}{$r_\mathrm{sP}$}&\multicolumn{2}{c}{}&-5.9\\
        \noalign{\smallskip}
        NIE & $23.3$&$1.3$& $0.4$&$0.6$ & $-0.1$&$0.6$ & $0.57$&$0.04$ & $-37.3$&$0.8$
& $11.1$&$1.2$ & \multicolumn{2}{c}{-}&\\
        \noalign{\smallskip}
        \bf{Galaxies} &
\multicolumn{2}{c}{$\sigma_\mathrm{DM}~(\mathrm{km}~\mathrm{s}^{-1})$} &
\multicolumn{8}{c}{}&\multicolumn{2}{c}{$r_\mathrm{cP}~(\mathrm{kpc})$}&\multicolumn{2}{c}{$r_\mathrm{tP}~(\mathrm{kpc})$}&\\
        \noalign{\smallskip}
        pJaffe-$L^*$ & $137$&$12$ & \multicolumn{2}{c}{-} & \multicolumn{2}{c}{-} &
\multicolumn{2}{c}{-} & \multicolumn{2}{c}{-} & \multicolumn{2}{c}{$[0.15]$} &
\multicolumn{2}{c}{$[45]$}&\\
\noalign{\smallskip}
     \bf{ICM}& \multicolumn{2}{c}{$\kappa_0$} & \multicolumn{8}{c}{}
&\multicolumn{2}{c}{$r_\mathrm{cP}$}&\multicolumn{2}{c}{$\beta$}\\
        \noalign{\smallskip}
        Main Clump & $0.091$&$0.015$&  \multicolumn{2}{c}{[3.7]} & \multicolumn{2}{c}{[8.7]} & \multicolumn{2}{c}{[0.39]} & \multicolumn{2}{c}{[-13]} & \multicolumn{2}{c}{[13.6]} & \multicolumn{2}{c}{[0.389]}& \\
        \noalign{\smallskip}
        Southern Tail & \multicolumn{2}{c}{[0.033]} & \multicolumn{2}{c}{[-23.4]} & \multicolumn{2}{c}{[-37.6]} & \multicolumn{2}{c}{[0.50]} & \multicolumn{2}{c}{[-37]} & \multicolumn{2}{c}{[170]} & \multicolumn{2}{c}{[1.8]}&\\ 
 
\noalign{\smallskip}
\hline

     \bf{Halo} & \multicolumn{2}{c}{$b~(\arcsec)$} & \multicolumn{8}{c}{}
&\multicolumn{2}{c}{$r_\mathrm{sP}$}&\multicolumn{2}{c}{}&3.2\\
        \noalign{\smallskip}
        NIE & $26$&$2$& $1.1$&$0.8$ & $-0.3$&$0.7$ & $0.53$&$0.04$ & $-39.0$&$1.1$& $12.9$&$1.0$ & \multicolumn{2}{c}{-}&\\
        \noalign{\smallskip}
        \bf{Galaxies} &
\multicolumn{2}{c}{$\sigma_\mathrm{DM}~(\mathrm{km}~\mathrm{s}^{-1})$} &
\multicolumn{8}{c}{}&\multicolumn{2}{c}{$r_\mathrm{cP}~(\mathrm{kpc})$}&\multicolumn{2}{c}{$r_\mathrm{tP}~(\mathrm{kpc})$}&\\
        \noalign{\smallskip}
        pJaffe-$L^*$ & $140$&$40$ & \multicolumn{2}{c}{-} & \multicolumn{2}{c}{-} &
\multicolumn{2}{c}{-} & \multicolumn{2}{c}{-} & \multicolumn{2}{c}{$[0.15]$} &
\multicolumn{2}{c}{$[45]$}&\\
pJaffe-BCG & $227$&$18$& \multicolumn{2}{c}{[0.0]} & \multicolumn{2}{c}{[0.0]} &
\multicolumn{2}{c}{[0.42]} & \multicolumn{2}{c}{[-32.8]} & \multicolumn{2}{c}{$[0.29]$} &
\multicolumn{2}{c}{$[87]$}&\\

\noalign{\smallskip}
     \bf{ICM}& \multicolumn{2}{c}{$\kappa_0$} & \multicolumn{8}{c}{}
&\multicolumn{2}{c}{$r_\mathrm{cP}$}&\multicolumn{2}{c}{$\beta$}\\
        \noalign{\smallskip}
        Main Clump & $0.093$&$0.012$&  \multicolumn{2}{c}{[3.7]} & \multicolumn{2}{c}{[8.7]} & \multicolumn{2}{c}{[0.39]} & \multicolumn{2}{c}{[-13]} & \multicolumn{2}{c}{[13.6]} & \multicolumn{2}{c}{[0.389]}& \\
        \noalign{\smallskip}
        Southern Tail & \multicolumn{2}{c}{[0.033]} & \multicolumn{2}{c}{[-23.4]} & \multicolumn{2}{c}{[-37.6]} & \multicolumn{2}{c}{[0.50]} & \multicolumn{2}{c}{[-37]} & \multicolumn{2}{c}{[170]} & \multicolumn{2}{c}{[1.8]}&\\ 
 
\noalign{\smallskip}
\hline
\end{tabular}
\caption{Model parameters inferred from the lensing analysis when the cluster-sized halo  is modelled as a NIE. $r_\mathrm{cP}$ is the projected core radius. The BCG is either scaled as the other galaxies  or varied independently (see models with a ``pJaffe-BCG" component). The ICM is either modelled apart (``Halo") or embedded in the cluster-sized halo (``Halo with ICM").}
\label{tab_best_is}
\end{table*}

In order to accomplish the strong lensing analysis, we performed a Bayesian investigation. The parameter probability distributions were then determined studying the posterior function. Computation of the likelihood function was based on the \texttt{gravlens} software \citep{kee01b,kee01a}. Such an analysis was performed in the source plane. Due to the large number of parameters and models, we exploited the Laplace approximation \citep{mac03}. The total number of constraints (42) is given by the coordinates of the observed image positions. The number of free parameters allowed to vary, i.e. the free parameters in the mass models plus the (10) unknown coordinates of the source positions, is 16 for a lens model with just a single cluster-sized halo. A further parameter may account for the velocity dispersion of the scaled galactic halos. As far as the BCG is concerned, we do not add parameters if the BCG is embedded in the cluster-sized halo or forced to follow the scaling law otherwise we add one further parameter if it is left free to vary. Finally, one further parameter accounts for the gas distribution when the ICM is modelled on its own. As a priori distribution for the parameters of the ICM, we consider the results from the X-ray analysis. 
%Finally, the number of degrees of freedom is given by the difference between the number of constraints and the number of free parameters.

In the present first attempt to include the ICM in a lensing analysis, we considered how and if the inclusion of gas improves the lensing modelling. An efficient way to compare different models exploits the Bayesian evidence $E$ \citep{mac03}. A difference of 2 for $\log E$ is regarded as positive evidence, and of 6 or more as strong evidence, against the model with the smaller value. 

Note that performing the fitting to just a single image system leads to a very small $\chi^2$-value for all the considered models independently of the system (A, E or S), since constraints associated to a single system are not enough to reliably determine the parameters. Only analysing all the image systems simultaneously leads to clear statements on the mass models.

\subsection{NFW profile}
\label{sec_nfwp}

We first considered NFW profiles. Models were then made of a cluster-sized NFW distribution and additional components for the galactic halos or the ICM, see Table~\ref{tab_best_fit}. Notation in Table~\ref{tab_best_fit} and in the following discussion distinguishes models according to the matter components included in the cluster-sized halo and to the modelling of the BCG. The NFW main halo describes the diluted DM plus some possible additional contributions. It can englobe either all the components at the same time (in the ``All'' model, DM and baryons are described by only one NFW profile), just the DM (``Halo''), DM and gas (``Halo with ICM''), DM and gas and BCG (``Halo with ICM and BCG'') or, finally, diluted DM plus all galactic halos (``Halo with all galaxies'') or plus just the BCG (``Halo with BCG''). When modelled apart, the galaxies can account for the BCG plus other ellipticals fitting a single scaling law (``All galaxies'') or just the other ellipticals without the BCG (``Other galaxies''). 

The simplest model is a single NFW halo, representing the total matter distribution (DM+ICM+galaxies). Its parameters are listed in Table~\ref{tab_best_fit}, see the ``All'' model. Even if the value of the scale length is larger than the range over which observational constraints are found, a combined fit to multiple source redshift image systems allows us to determine $r_\mathrm{sP}$ and its uncertainty \citep{lim+al08}. This is crucial in the estimate of the concentration, see Sec.~\ref{sec_conc}. With this simple mass model we were able to reproduce the observed images much better than assuming a single isothermal profile, see Sec.~\ref{sec_isot}. All the images were reproduced with a mean distance of $\simeq 0.6\arcsec$. Since all the priors on the parameters are flat, it makes sense to consider the $\chi^2(\simeq 32.1)$ of the inferred model. Even for a quite complex system as AC~114 a single NFW model, accounting at the same time for dark matter, stars and gas, can provide a good fit to the data with a reduced $\chi^2_\mathrm{red}\simeq1.2$.

The subsequent addition of ICM and galaxy-sized halos considerably improved the fit, but above all helped to achieve a physically more consistent model, which better describes the features of the cluster. As a first step, we followed the usual approach and considered galactic halos  together with a cluster-sized component. For such models (``Halo with ICM'' and ``Halo with ICM and BCG'' in Table~\ref{tab_best_fit}), all the diluted mass distributions, i.e dark matter plus gas, contribute to a single cluster-sized NFW halo. The fitted parameters for each component are listed in Table~\ref{tab_best_fit}. Letting the BCG be free to vary, the degeneracy between the cusped cluster-sized halo and the BCG halo takes over and the posterior probability is maximum for a cD galaxy with null mass. We then limited our analysis to a BCG halo either following galactic scaling laws (``Halo with ICM'' and ``All galaxies'') or embedded in the cluster-sized one (``Halo with ICM and BCG'' and ``other galaxies''). In both cases, the evidence is much better than for a single NFW halo. Note that the listed values of the evidence are given apart from a constant factor depending on the data and a second hidden factor depending on the flat priors on the parameters of the NFW profile, which is constant across the models.

As a second step, we considered the effect of explicitly modelling the gas distribution. For all the analysed models, adding a component for the ICM improves the evidence. In Bayesian analysis, given equal priors for the different hypotheses, model are ranked by evaluating the evidence. Then, on a statistical point of view it is better to model the gas independently from the DM cluster-sized halo. The physical reason beyond that is that the ICM does not follow the mass. In a relaxed cluster, the gas follows the gravitational potential and is rounder than the mass distribution. AC~114 is dynamically active and our analysis shows that the differences between gas and dark matter distribution are further exacerbated.

Models whose cluster-sized halo has to account for either DM+gas+galaxies (``All'') or DM+BCG+gas with other galaxies modelled separately (``Halo with ICM and BCG'' and  ``Other galaxies'') or DM+galaxies with gas modelled separately ( ``Halo with all galaxies'' with ``ICM'') or DM+BCG with other galaxies and gas modelled separately (``Halo with BCG'' and ``other galaxies '' and ``ICM'') have an evidence of $\log E \sim -12.4$, $-7.6$ and $-4.0$ or $0.0$, respectively. Adding physical motivated components increases the evidence. Such a trend is also confirmed by a different version, the model where the cluster-sized components accounts only for the diluted DM  whereas the BCG follows the galactic scaling laws and the ICM is modelled separately (``Halo'' and ``Galaxies with BCG'' and ``ICM''). The corresponding evidence ($\log E \sim -1.9$)  is better than those of models without either the galaxies or the gas. This confirms that the results is independent on the modelling of the BCG.

Since accounting for the gas is quite unusual in lensing analyses, let us compare the models accounting for a gas component with the usual way, i.e. a composite mass distribution in which both ICM and DM are parameterized altogether as a single NFW profile. Such usual model provides a good fit to the data  either for the BCG scaled together with the other galaxies (``Halo with ICM'' and ``All galaxies'',  $\chi^2\simeq 24.8$) or for the BCG  embedded in the cluster-sized halo (``Halo with ICM and BCG'' and ``Other galaxies'',  $\chi^2\simeq 21.8$). Note that the inclusion of both galactic and ICM components is needed to improve the fit, whereas accounting only for the gas is not helpful. It is the physical information drawn from X-ray data that demands for the inclusion of the ICM in the modelling. Our analysis shows then that adding physical motivated complexity to the lensing models (either in the form of galactic halos or in the form of diluted gas distribution) improves the description of a cluster lens both under the statistical and the physical point of view.

Note that as far as a simple $\chi^2$-analysis goes on, adding the gas component would not be justified for AC~114, since the fit is not significantly improved. This point needs to be further investigated considering a sample of clusters with different X-ray surface brightness slopes.

Modelling the gas helps to better investigate the DM halo. Comparing the properties of the cluster-sized DM halos (with or without galaxies), see Table~\ref{tab_best_fit}, to DM+gas halos (with or without galaxies), two properties stand out. First, in order to account for the mass contributed by the gas, the DM+ICM halo has larger central convergence and larger radius. The two parameters varies accordingly in such a way to leave the concentration nearly unchanged. Second, due to the misalignment between gas and DM, the DM+gas halo turns out to be rotated counter-clockwise with respect to the only DM component.

The addition of the ICM, which is quite flattened, caused a slight decrease of the projected scale length, $r_\mathrm{sP}$, for the DM component, whose orientation experienced a clock-wise rotation of $\sim 2\deg$. These changes are slight, as the ICM has a relative low mass compared to the dark matter, but nevertheless interesting. The decrease of $r_\mathrm{sP}$ shows that the dark matter component is more compact than the ICM. The total projected mass within $75~\mathrm{kpc}$ ($150~\mathrm{kpc}$) is $[3.8\pm 0.4]\times 10^{13}M_\odot$ ($[11.3\pm 1.0]\times 10^{13}M_\odot$), in good agreement with previous estimates \citep{nat+al98,def+al04}.

Ellipticity and orientation of the dark matter component are almost the same as the ones of the southern component of the ICM, whereas the northern component, which is the main baryonic component in the cluster core, is less elliptical and rotated counter-clockwise of $\sim 27\deg$ compared to the dark matter. Its centroid is displaced of $\sim 40~\mathrm{kpc}$ from the center of the dark matter distribution. This evident spatial offset between the dark matter and the main baryonic component in the cluster core brings evidence that the cluster is not in equilibrium. The fact that between the center of the dark matter component and the position of the BCG there is no significant offset portends that the dark matter behaves like collisionless particles during the merging process.

We can test the predictive power of our model by guessing the source redshifts of the multiple image systems without spectroscopic confirmation. Both the B and D systems are strongly perturbed by local galaxies and a prediction would require a detailed modelling of galactic halos, which is beyond the scope of our analysis. The system C is not affected by such a problem. Such a three image system has a predicted lensing redshift of $\sim 2.3$ in agreement with \citet{cam+al01}.

%Some further tests on the model with a DM+ICM cluster-sized halo plus galactic components allowed us to asses which galaxies have the major impact on the fitting procedure. It turns out the the bulk of their effect is due to the four most massive galaxies. When including them in the modelling, we could get a $\chi^2 \ls 28$. On the other hand, when excising the six less massive galaxies from the analysis, the $\chi^2$ is nearly unchanged.

\subsection{Isothermal profile}
\label{sec_isot}

Alternatively to the NFW model, we considered an isothermal profile for the main mass component, see Table~\ref{tab_best_is}. When the BCG is modelled apart from the other galaxies an additional entry line shows up (``pJaffe-BCG''); in absence of such a line, the BCG halo follows the standard scaling law for ellipticals. As a first step, we modelled AC~114 with a single NIE, representing all the matter present in the galaxy cluster. As in the NFW parameters we used flat priors. This model, which turned out to be centred in the neighbourhood of the BCG galaxy, was quite inadequate. The reason is that the central density of this model is too low and therefore the central caustic too narrow, which in turn implies the vanishing of the merging images A4 and A5. 

To solve this issue we added the mass distribution from galaxy-sized halos. Differently from the cusped NFW, a cored NIE needs an additional peaked mass distribution associated with the BCG to provide good fit to the data. Due to the degeneracy between BCG and cluster-sized DM halo is then misleading to interpret the DM distributions studied in this section as pure cored isothermal ones. Due to the small core radius imposed on the BCG, the overall profile has an effective central divergence which comes afloat from the cored NIE. The distribution is isothermal, $\rho \simeq r^{-2}$, only at large radii ($\gs 12\arcsec$). Note that the fit, and consequently the evidence, improves significantly when the BCG is not forced to follow the galactic scaling laws.

We finally considered separately the ICM, the galaxy sized halos and the dark matter component modelled as a NIE profile. The total projected mass within circles with radii $75~\mathrm{kpc}$ (150 $\mathrm{kpc}$) is $[4.0\pm 0.3]\times 10^{13}M_\odot$ ($[11.2\pm 0.9]\times 10^{13}M_\odot$), in agreement with the estimate based on the assumption of DM distributed as a NFW profile and with previous results in \citet{cam+al01}, who used a slightly different modelling, i.e pseudo-Jaffe profiles for both galactic and cluster-sized halos. We remark that they had to consider additional NW and SE sub-structures to account for weak lensing effects outside of the very inner core.

As in the NFW case, the addition of the ICM mass components was not able to significantly improve the $\chi^2$, since its mass distribution is widely distributed, with a subcritical surface density which is unable to produce any strong lensing. Only its total mass has an influence on the lensing properties of the cluster.  On the other hand, the evidence factor increases for each subsequent addition of physically motivated components. Models with an explicit component for the ICM have larger evidences than corresponding models without. This result is then independent of the parameterization of the DM halo.

The value of $\sigma_\mathrm{DM}$ for the $L^*$ galaxy has to be much higher assuming an isothermal profile for the DM than for a NFW distribution. In fact, the cored NIE is quite inadequate as a model for the DM so that galaxies, and in particular the BCG, have to supply additional convergence to broaden the central caustic. The discrepancy between the values of $L^*$ inferred under different hypotheses, see Table~\ref{tab_best_fit} and Table~\ref{tab_best_is}, gives an estimate of the systematic uncertainties that plague galactic parameters inferred from lensing in our analysys.

%A main prediction of $N$-body simulations is that dark matter halos have a universal profile. NFW profiles are strongly favoured over isothermal models. On the scale of galaxy clusters, lensing observations are confirming such a view. Stacking weak lensing clusters, \citet{oka+al09} found that the isothermal profile is highly disfavoured with respect to the NFW model and that the measured profile at small radii is consistent with the inner NFW slope. An additional hint in this direction is also given by studies of arc magnification \citep{shu+al08}.

%Both parametric and non parametric studies of other strong lensing clusters, which alike to AC~114 presents sets of multiple lensed sources at different redshifts, also support a universal profile ($\alpha \sim 1$) in the inner regions \citep{sah+al06,lim+al08,sa+re09}. In fact, multiple lensed sources at different redshifts can break the degeneracy between the scale radius and the inner slope that can otherwise plague strong lensing analyses \citep{san+al08}. Even when the preferred value for $r_\mathrm{s}$ is found larger than the clustercentric distance of images, lensed sources at different redshifts are actually sensitive to the scale radius \citep{lim+al08}. 

%Our analysis of AC~114 seems to suggest that on cluster-scales a single isothermal profile can not provide an adequate description of the mass distribution. 

Whereas a single NIE is unable to provide a good fit, once a second central peak associated with the BCG is added, the overall profile is no more isothermal. On the other hand a single NFW profile provides a good description for the overall mass distribution. However such an advantage gets lost when we focus on the cluster-sized dark matter distribution instead of the overall distribution. As far as we add separate components for the gas and the galactic halos, either an isothermal or a NFW profile for the dark matter give a good fit. Exploiting evidence should be the way to compare the two scenarios, but we are cautious to do it here for two main reasons. First, the evidence factors were computed apart from a factor with depends on the priors on the cluster-sized halo parameters. Since we considered flat priors, the evidence depends on the allowed range. This a priori factor has no effect on the model comparison given a shape for the halo, but could affect the comparison between the isothermal and the NFW profile.

Furthermore, due to the large number of models and discontinuities, mainly associated with the central radial caustic, we had to perform our analysis in the source plane. Computation of the likelihood in the image plane for a number of models showed that the $\chi^2$ values in the source plane may be overestimated  by $\gs 2$ for the models with a NFW components, whereas are underestimated by $\gs 2$ for the isothermal case.

Such effects can considerably affect any model comparison, so we prefer to address this problem in a future work investigating a larger sample of clusters.

\section{Results}
\label{sec_resu}

Let us review the results that directly follow from our multi-component parametric approach. 

\subsection{Dynamical status}

The strong lensing analysis of the inner regions of AC~114 brings new evidence about its dynamical status. We attempted a new multi-wavelength approach in which the baryonic components were mainly constrained using observations either in the X-ray or optical band, allowing us to infer directly the dark matter distribution from the lensing analysis. The gas is displaced from the dark matter. The main X-ray clump and the cluster-sized DM halo are off-centered by $\sim 9\arcsec$, an offset much larger than the Chandra accuracy of $\sim 1\arcsec$ which determines the accuracy in the X-ray peak position. The relative orientation differs by $(27\pm 4)\deg$. On the other hand, the DM clump is nearly aligned with the X-ray tail. This provides further evidence that the X-ray surface brightness in the core is strongly perturbed by the dynamical activity. The likely motion of a sub-structure toward north-east, as suggested by the fronts, might have distorted the local emission causing a rotation of the overall surface brightness of the central X-ray clump towards East and the relative misalignment of gas and dark matter.

Note that the above analysis is limited to the very inner regions probed by strong lensing. When averaged over larger scales, distribution features might differ and the impact of substructure should be properly addressed.

\subsection{Collisionless dark matter}

Whereas the ICM is clearly displaced, the dark matter distribution follows the galaxy density. The quite large errors in the parameters describing either the number, the luminosity or the stellar mass density distributions, see Table~\ref{tab_gal_den}, make it difficult to distinguish if a certain galaxy density traces the dark matter distribution better than other ones. However, the good agreement between each other allows us to draw some conclusions. The galaxy and dark matter distributions share comparable centroid position, orientation and ellipticity. Since dark matter was modelled with a cusped profile whereas the galaxy density were fitted to a cored distribution, the comparison can not be extended to the remaining parameters. The agreement is a further hint on the collisionless nature of dark matter, as suggested from the bullet cluster \citep{clo+al04}. This time, we could probe that the agreement between galaxies and dark matter concerns not only the location but also the shape of the distribution.

When comparing dark matter with the galaxy distributions the agreement becomes striking when we consider the number density distribution. As written before, errors are quite large and definite statements can not be drawn but the similarities between the expected values are nevertheless noteworthy. Galaxy abundance has been considered as a proxy for the cluster mass \citep{hic+al06}. Our result might provide further indication that galaxy number density is a dependable tracer also for DM shape and orientation.

\subsection{Baryons and dark matter}

\begin{figure}
        \resizebox{\hsize}{!}{\includegraphics{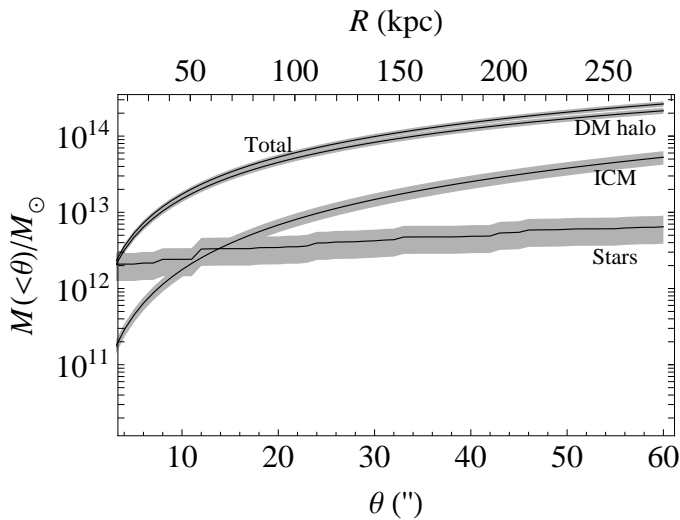}}
        \caption{Mass, in units of $M_\odot$, enclosed within a given projected radius for each component. Total mass, DM halo, ICM mass and the stellar contribution from galaxies are plotted from the top to the bottom. DM halo refers to the cluster-sized dark matter component found in the multi-wavelength lensing analysis.}
	\label{fig_mass_ins}
\end{figure}

\begin{figure}
        \resizebox{\hsize}{!}{\includegraphics{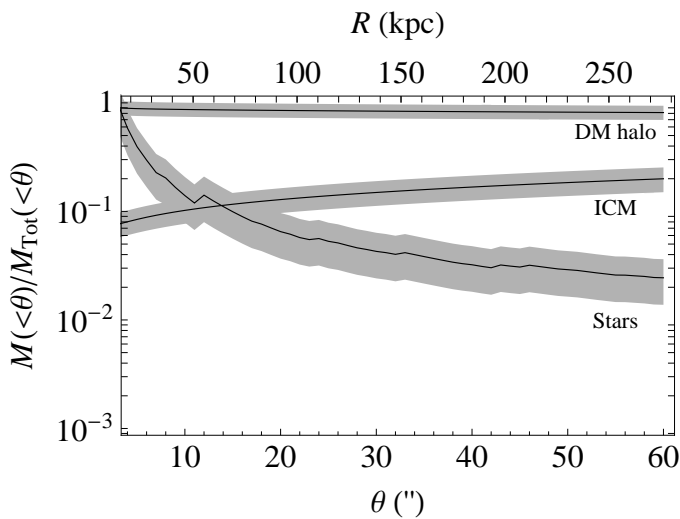}}
        \caption{Mass fractions as a function of the projected radius. Notation is the same as in Fig. \ref{fig_mass_ins}.}
	\label{fig_frac_mass_ins}
\end{figure}

Our multi-wavelength approach allows us to determine the mass profile of each component in the very inner regions. Figures~\ref{fig_mass_ins} and \ref{fig_frac_mass_ins} show the enclosed projected masses for clustercentric distances less than $1\arcmin$ ($R \ls 280~\mathrm{kpc}$). We consider the two main baryonic components (stars in galaxies and hot ICM), the cluster-sized dark matter halo and the total projected mass (as modelled with a single NFW profile, see Sec.~\ref{sec_nfwp}). The mass values with the smaller errors are those from the lensing analysis. The estimates of the different mass components scale differently with the Hubble constant, so that for comparison we fixed $h=0.7$. Note that we consider projected mass distributions for each component, which avoids biases due to comparing projected with three-dimensional quantities. We remind that the mass of each component has been derived with a different method: the DM distribution has been inferred from lensing whereas the ICM and the stellar mass were estimated from X-ray, see Sec.~\ref{sec_xray}, and optical light data, see Sec.~\ref{sec_stel}, respectively.

Typical trends are retrieved \citep{bi+sa06}. The dark matter halo is the dominant component ($\sim 80\pm 10\%$ at $R \sim 280~\mathrm{kpc}$). In the center ($R \ls 50~\mathrm{kpc}$) the baryonic budget is dominated by the stellar mass in the BCG, whereas the ICM contribution takes over at larger radii. The gas distribution is shallower than the dark matter profile, so that the ICM fraction increases from $\sim 10\pm 3\%$ at $R \sim 50~\mathrm{kpc}$ to $\sim 20\pm 5\%$ at $R \sim 280~\mathrm{kpc}$. These values are larger but still compatible with typical values inferred with X-ray analyses of luminous clusters \citep{all+al08}. On the other hand, the stellar fraction is just a few percents at $R \gs 150~\mathrm{kpc}$ \citep{bi+sa06}.

The luminosity function of AC~114 has been extensively studied \citep{and+al05}. Adopting a total luminosity of $(1.5 \pm 0.2)\times 10^{12} L_\odot$ in $r$ and $(1.9 \pm 1.2)\times 10^{11} L_\odot$ in $B$ within $0.6~\mathrm{Mpc}/h$ \citep{kr+be07}, we get mass-to-light ratios of $M/L_r =(670\pm 110) M_\odot/L_\odot$ and $M/L_B =(4400\pm 1400) M_\odot/L_\odot$, which point to a underluminous cluster core. However, due to the large errors, especially in the $B$-band, the mass-to-light ratios are still slightly compatible with estimates from other clusters \citep{rin+al04,biv08}.

\section{Comparison with theoretical predictions}
\label{sec_comp}

It can be of interest to compare the results of our analysis with expectations from $N$-body simulations or theoretical studies. The comparison requires an extrapolation of our mass model to larger radii, well beyond what is directly probed by strong lensing. For this reason, we decided to limit the analysis in this section only to the global fits, i.e. to mass models which describe the total matter distribution (baryons and dark matter) with a single cluster-sized halo. The following considerations can then be seen as a corollary to the main result of the paper, i.e. that ICM can matter in lens modelling, and do not make use of the explicit gas modelling discussed before.

%\subsection{Universal vs isothermal profiles}
%\label{sec_univ}

\subsection{Inner slope}
\label{sec_inne}

Values of inner and outer slopes of density profiles coming out from $N$-body simulations are still debated with different parameterizations competing \citep{mer+al06,sa+re09}. Baryons play a role too, since their infalling would steepen the dark matter profile. However, in large clusters this effect is expected to be small exterior to $\sim 20~\mathrm{kpc}$ \citep{gne+al04}. The general consensus is that in the inner regions of clusters the dark matter profile should go as $\rho \sim r^{-\alpha}$, with $\alpha$ between $1$ and $1.4$ \citep{die+al04}. 

In order to investigate the inner slope, we considered a total matter distribution modelled as a singular softened power law  ($r_\mathrm{c}=0$), which represents a power law mass profile, i.e. $\rho\propto r^{-\alpha}$. This mass profile is able to reproduce all the images of the observed systems, and for the slope we obtained a best fit value of $\alpha=1.38\pm0.01$. Such estimate is mainly due to the images near the central radial caustic.

The simple power-law used in our analysis make very prompt the comparison with previous analyses which employed the same parameterization, showing good agreement \citep{sah+al06}. On the other hand, the very small uncertainty on $\alpha$ is more due to not enough accurate modelling than to very precise statistical accuracy. Since we were mainly interested in comparison with $N$-body simulations, we modelled the total mass profile as a single power-law. We did not attempt to distinguish a baryonic from a DM component. Inferring the inner slope of the DM component would require a much more detailed modelling \citep{lim+al08}. Different parameterizations of the cD galaxies can bring about an uncertainty of $\sim 0.05$ on the inner slope \citep{lim+al08}. On the other hand, as far as the ICM mass distribution is modelled with a cored profile, the estimate of $\alpha$ does not depend on the inclusion of the gas in the fit procedure.

The main source of error ($\sim 0.1$) is due to the absence of a length scale in the simple power-law profile we used. We can quantify the uncertainty with the following simple reasoning. The slope of a NFW profile changes from $\alpha =1$ in the very inner regions to $\alpha \simeq 1.18$ at $r \simeq 10^{-1}r_\mathrm{s}$, with a mean value of $\langle \alpha\rangle \simeq 1.12$. Then, for sets of multiple images covering nearly one tenth of the length scale, modelling the profile with a power-law causes an over-estimate of the inner slope $\alpha$ of $\sim 0.1$. A further source of error is due to the degeneracy between the slope and the scale radius of a generalized NFW profile. \citet{lim+al08} showed that fixing $r_\mathrm{s}$ to a value smaller that the best fit estimate causes an underestimate of the slope. The related uncertainty is $\sim 0.05$ \citep{lim+al08}. Even after accounting for such systematics, we see that the estimated value of the inner slope of AC~114 is still steeper than a simple NFW profile and falls just in the middle of the range compatible with theoretical predictions \citep{die+al04}.

\subsection{Concentration}
\label{sec_conc}

The concentration parameter reflects the central density of the halo, bringing imprints of the halo assembly history and thereby of its time of formation. Dealing with ellipsoidal halos, we need generalized definitions for the intrinsic NFW parameters. We follow \citet{co+ki07}, who defined a triaxial radius $r_{200}$ such that the mean density contained within an ellipsoid of semi-major axis $r_{200}$ is $200$ times the critical density at the halo redshift; the corresponding concentration is $c_{200} \equiv r_{200}/ r_\mathrm{s}$. Then, the characteristic overdensity in terms of $c_{200}$ is the same as for a spherical profile. The mass, $M_{200}$, is the mass within the ellipsoid of semi-major axis $r_{200}$. Such defined $c_{200}$ and $M_{200}$ have small deviations with respect to the parameters computed by fitting spherically averaged density profiles, as done in  $N$-body simulations. The only caveat is that the spherical mass obtained in simulations is significantly less than the ellipsoidal $M_{200}$ for extreme axial ratios \citep{co+ki07}.

%If a cluster is elongated along the line of sight, the concentration parameter and the virial mass estimated from lensing are overestimated \citep{gav05,ogu+al05}. On the other hand, there are more inefficient lensing orientations for a triaxial halo than there are efficient ones \citep{cor+al09}. Investigations in the weak lensing regime demonstrated that neglecting halo triaxiality can lead to over- and under-estimates of up to 50\% and a factor of 2 in halo mass and concentration, respectively \citep{co+ki07}. Even assuming statistical priors on the intrinsic shape, uncertainties are still large \citep{cor+al09}. 

We estimated $M_{200}$ and $c_{200}$ from the projected NFW parameters directly inferred from the fit, i.e. $r_\mathrm{sP}$ and $k_\mathrm{NFW}$, see App.~\ref{app_proj}. Let us consider the single NFW profile accounting for the overall mass distribution, see Sec.~\ref{sec_nfwp}. A main source of uncertainty is due to triaxiality issues  \citep{gav05,ogu+al05,cor+al09}. A simple way to account for projection effects is detailed in App.~\ref{app_proj}. The expected values of the geometrical correction factors can be estimated assuming random orientations and intrinsic axial ratios with probability density following results from $N$-body simulations \citep{ji+su02}. In order to compare our results with theoretical predictions, we did not consider a generalized NFW profile but we fixed the inner slope to $\alpha=1$, see Sec.~\ref{sec_nfwp}. We obtained $c_{200}= 3.5 \pm 0.7$, in good  agreement with the estimate based on the velocity dispersion derived in Sec.~ \ref{sec_viri}, and  $M_{200}=(1.3 \pm 0.9)\times 10^{15}M_\odot/h$, slightly lower than the mass from the virial theorem, see Sec.~ \ref{sec_viri}. Note that neglecting projection effects, the error on $c_{200}$ would have been $\sim 0.2$. We get nearly the same value for $c_{200}(=3.7\pm 0.8)$ if we consider the cluster-sized DM halo instead of the total mass distribution. Due to the flatness of the ICM distribution, the central value for the concentration of the DM halo is higher than considering the overall distribution. However, the shift is smaller that the statistical uncertainty

The halo concentration parameter is expected to be related to its virial mass, with the concentration decreasing gradually with mass \citep{bul+al01}. According to recent numerical simulations \citep{duf+al08}, the concentration of a cluster with the same mass just derived for AC~114 at its redshift should be $c_{200}= 2.90 \pm 0.13 \pm0.13$. The agreement with our result is striking, something very unusual when comparing concentrations derived from lensing analyses to predicted values.

\section{Discussion}
\label{sec_disc}

Exploiting in a combined way lensing observations with multi-wavelength data sets is a very powerful tool to constrain properties of galaxy clusters \citep{fo+pe02,clo+al04,smi+al05,ser07,lem+al08}. Here, we have performed a lensing analysis in which the gas mass distribution, previously inferred from X-ray observations, has been embedded from the very beginning in the modelling. Gas is the main baryonic component and typically contributes for $\gs 10\%$ of the total mass in galaxy clusters \citep{all+al08}. Considering the ICM in the parameterization can be see as an improvement with respect to the usual way of modelling only cluster-sized dark matter and galaxy-sized halos. We attempted a first step in this direction.

The main result of this paper is that the explicit inclusion of the gas distribution plays a role in lensing modelling. Whereas is well known that galactic halos have to be added to increase the accuracy in strong lensing modelling, we showed that the inclusion of an additional component accounting for the gas distribution helps too. Models in which the cluster-sized dark matter halo is considered together with the ICM perform better than parametric solutions with a single halo accounting at the same time for both dark matter and diluted baryons. The physical gain is quite significant too, since modelling the gas on its own allows us to put observational constraints directly on the dark matter distribution. This is crucial in the understanding of the formation of cosmic structures and the co-evolution of baryons and dark matter in clusters of galaxies.

X-ray data are usually exploited to investigate the mass distribution of a galaxy cluster on a larger scale than the very inner regions analysed with strong lensing analyses. However, X-ray telescopes map the intracluster medium in the central parts too. Our method aims to exploit such information. We do not rely on usual hypotheses needed to infer the total mass from X-ray data (hydrostatic equilibrium, constant baryonic fraction, ...) which are better satisfied within a radius as large as the virial one. We just use the the gas distribution as directly measured from X-ray observations, which is reliable on each scale. As far as a study of the inner regions is concerned no calibration on a larger scale, as those provided by weak-lensing studies, is then needed.

As test-bed we considered AC~114, a dynamically active cluster. Comparison of the dark matter map directly obtained from lensing modelling with either the gas or the stellar mass distribution can give a deep insight on the properties of the cluster. Our analysis of AC~114 provided a further example that the ICM is displaced from the dark matter in dynamically active clusters, whereas the collisionless nature of dark matter is probed by the good matching with the galaxy distribution. 

Our lensing analyses left us with a mass modelling of AC~114, which can be compared with $N$-body simulations. The obtained results are in remarkable agreement with predictions. For AC~114, we found that: $i)$ a cusped NFW model for the overall mass distribution seems to be preferred over an isothermal profile; $ii)$ the inner slope is slightly steeper than a simple NFW; $iii)$ the concentration parameter is in line with predictions from mass-concentration scaling relations.

However, comparisons with $N$-body simulations must be taken cum grano salis. Statistical samples of halos with the mass of a galaxy cluster are very demanding to obtain with numerical simulations. On the other hand, AC~114 is dynamically active, which makes the comparison even more ambiguous. Furthermore, our estimation of the viral mass and of the concentration parameter required an extrapolation to scales much larger than what mapped by strong lensing. Calibration with other methods, such as weak lensing, is then needed to support conclusions on the global properties. Nevertheless, we feel encouraged by the agreement between the parameters estimated with lensing and those inferred with a dynamical analysis.

Our estimated inner slope is in agreement with estimate from numerical simulations. A value of $\alpha \gs 1$ might be indication of steepening due to adiabatic contraction, but the very young dynamical age of AC~114 and its intense ongoing merging activity weaken such an interpretation. It is noteworthy that the recent modelling of the seemingly relaxed cluster A~1703 preferred an inner slope larger than one ($\alpha \sim 1.1$) too \citep{lim+al08}.

Different models used for the BCG do not change our results. For a comparison with numerical simulations, we used global fits (i.e. a single DM+baryon halo); as far as comparison between DM and gas (or light) is concerned, different assumptions on the BCG do not affect significantly either orientation or centroid of the DM distribution. Finally as far as the integrated mass distributions of the different components are concerned, the stellar mass estimate is based solely on measured photometry wheraes the gas mass is derived from X-ray observations.

Lensing clusters appear to be quite over-concentrated \citep{co+na07,joh+al07,bro+al08,man+al08,og+bl09,ogu+al09,oka+al09,cor+al09}. The analysis we performed provides some new elements. First, some peculiarities might make our results less affected from biases. We derived the concentration parameter using only strong lensing data and we did not use the spherical approximation for the halo profile. In fact, different definitions of parameters for spherically averaged profiles can play a role when comparing observations to predictions \citep{br+ba08}. Second, AC~114 has some peculiar features that might make the high concentration problem much less pronounced. In particular, the very long tail in the X-ray morphology and the detection of a shock front suggest that the cluster develops in the plane of sky. The elongation of the cluster could be probed observationally combining lensing and X-ray data with measurements of the Sunyaev-Zeldovich effect \citep{fo+pe02,def+al05,ser+al06,ser07}. Unfortunately, detection for AC~114 is still marginal \citep{and+al96} and deeper radio observations are needed.

\begin{acknowledgements}
     The authors thank E.~De Filippis and P.~Martini for some useful clarifications. For the first stages of this work, M.S. has been supported by the Swiss National Science Foundation.
\end{acknowledgements}

\appendix

\section{Velocity dispersion}
\label{app_vel}

Cluster galaxy velocity dispersion is a crucial source of information. We collected positions and redshifts of galaxies in the vicinity of AC~114 from the NASA/IPAC Extragalactic Database (NED). We retrieved 248 galaxies within $\ls 1.5~\mathrm{Mpc}/h$ from the BCG in the redshift range $0.1 \le z \le 0.5$. A careful treatment of interlopers is required in dynamical modelling. Many approaches have been proposed and their efficiency has been tested using numerical simulations \citep{woj+al07,biv+al06}. Here, we propose a method of interloper removal which combines several of them.

In order to select member galaxies, we first exploit velocity information using an adaptive kernel technique \citep{pis93,pis96}. Such a nonparametric method evaluates the underlying density probability function from the observed discrete data-set. We identify the main peak in the distribution and reject galaxies not belonging to this peak. This cut has been successfully employed a number of times \citep{gir+al98,gi+me01}. To evaluate the optimal smoothing parameter, we minimise the integrated square error \citep{pis93} fixing the initial value to the estimate proposed by \citet{vio+al94}. As a second step, we take into account both the position and the velocity information by using the procedure of the shifting gapper \citep{fad+al96,gir+al98}. Such a method combine velocity information with the clustercentric radial distance. In each bin, shifting along the radial distance form the centre, a galaxy is removed if separated from the main local body by more than a fixed gap in velocity. We use a gap of $\ge 1000~\mathrm{km~s}^{-1}$ in the cluster rest frame and a bin of $0.3/h~\mathrm{Mpc}$.

As a third and final step, we employ a Bayesian technique \citep{and+al08}. The effect of a contaminating population can be inferred by considering that data $v_i$ come from a Gaussian-distributed intensity super-imposed on an homogeneous random process   \citep{mar+al00,ma+ge04,and+al08},
\beq
p(v_i; f_\mathrm{cl},v_\mathrm{cl},\sigma_\mathrm{los}, \Delta v )=f_\mathrm{cl} {\cal{N}}(v_\mathrm{cl},\sigma_\mathrm{los})+(1-f_\mathrm{cl})/\Delta v,
\eeq
where $f_\mathrm{cl}$ is the fraction of member galaxies, $ {\cal{N}}$ is a normal distribution centered on $v_\mathrm{cl}$ and with dispersion $\sigma_\mathrm{los}$ and $\Delta v$ is the velocity range spanned by data. The likelihood is then
\beq
{\cal{L}} \propto \prod_i p(v_i;...) .
\eeq
After considering a sharp prior on $\Delta v= \max \{ v_i\} -\min \{ v_i\}$ and flat priors on the other parameters, we get the final probability. Such a powerful statistical method, which guesses the total fraction of interlopers without actually picking them out, has been employed to constrain the phase-space probability function \citep{mar+al00,ma+ge04}, but is rarely used to constrain the velocity dispersion \citep{and+al08}. After marginalization, we obtain estimates of $z_\mathrm{cl}=0.3153 \pm 0.0007$ for the cluster mean redshift and $\sigma_\mathrm{los} = 1900 \pm 100~\mathrm{km~s}^{-1}$ for the velocity dispersion, the estimated fraction of interlopers being $(11 \pm 6)\%$, in good agreement with the estimates from numerical simulations that showed that using non-Bayesian methods $\sim 18\%$ of selected members are unrecognized interlopers \citep{biv+al06}. Standard corrections for cosmological effects and velocity errors ($\delta v \sim150~\mathrm{km~s}^{-1}$) have been applied \citep{dan+al80}. Our final estimates of $z$ and $\sigma_\mathrm{los}$ are remarkably stable for different thresholds and gaps in the first steps of our selection procedure. Only the estimated fraction of non members is sensitive to the details of the previous cuts. The results of the Bayesian technique just employed are also stable in the case of an intrinsically skewed velocity distribution \citep{and+al08}.

Our expected value for $\sigma_\mathrm{los}$ is lower, but compatible within errors, than the recent estimate from \citet{mar+al07}, but higher, despite still marginally compatible, with early estimates from \citet{co+sh87} and \citet{ma+ge01}. On the other hand, the estimate from \citet{gi+me01}, who considered only a sample of non active galaxies, is quite lower.

\section{Substructures}
\label{app_subs}

Substructures in the galaxy distributions of AC~114 have been detected with several tecniques. Two clumps of galaxies, the first one northwest (NW clump) and the second one southest (SE clump) of the BCG were noted in \citet{nat+al98}, who combined optical and weak lensing data. According to a combined weak and strong lensing modelling \citep{cam+al01}, the NW clump at $\{ \theta_1, \theta_2\}\sim\ \{80\arcsec, 30\arcsec \}$ and the SE clump at $\{-75\arcsec, -75\arcsec\}$ have a mass $\sim 20\%$ and  $\sim 35\%$, respectively, of the mass of the main clump associated with the central cD galaxy.  A clump of galaxies looking like a group with its own cD-like galaxy at $\sim 1.1~\mathrm{Mpc}/h$ northwest of the BCG was noted in \citet{kr+be07}, who also found evidence for associated intracluster light emission. Actually, the main galaxy in the NW clump is as luminous as the central BCG galaxy \citep{cou+al98,sta+al02}. The luminosity map reveals also other features, most notably a filament towards northest in the very inner region, nearly perpendicular to the overall orientation.

Combined galaxy velocity and position information can single out local substructures and compact subsystems. Here, we want to exploit such information. The $\Delta_\mathrm{DS}$ test \citep{dr+sh88} looks for significant deviations in local groups that either have an average velocity $\bar{v}_\mathrm{loc}$ that differs from the cluster mean, $\bar{v}_\mathrm{glo}$, or have a velocity dispersion, $\sigma_\mathrm{loc}$, that differs from the global one, $\sigma_\mathrm{glo}$. For our analysis, we considered the classic version of the test, which considers all possible subgroups of ten neighbors around each cluster galaxy \citep{dr+sh88} with the only slight difference that for calculating location and scale we consider the biweight estimators instead of mean and standard deviation \citep{bee+al90}. Then, the deviation for each galaxy can be expressed as \citep{dr+sh88} 
\beq
\delta_\mathrm{DS}^2=	\frac{11}{\sigma_\mathrm{glo}^2}\left[ (\bar{v}_\mathrm{loc}-\bar{v}_\mathrm{glo})^2+ (\sigma_\mathrm{loc}-\sigma_\mathrm{glo})^2\right].
\eeq 
The parameter $\Delta_\mathrm{DS}=\sum \delta_\mathrm{DS}$ quantifies the overall presence of substructures. The $\Delta_\mathrm{DS}$ statistic is then obtained by comparing the measured value with those obtained from simulated samples generated randomly shuffling velocities. The probability $P_{\Delta_\mathrm{DS}}$ that the observed value is due to noise is then given by the fraction of samples with a value of $\Delta_\mathrm{DS}$ larger than the observed one. We get  $P_{\Delta_\mathrm{DS}} \ls 0.1$. We also considered modified versions of the test \citep{biv+al02}. Results are not affected. 

The scope of the method can also be adapted to find which galaxies have the highest likelihood of residing in subclusters. This can be done considering the $\delta_\mathrm{DS}$-statistic in a similar way to what just done for $\Delta_\mathrm{DS}$. The $\delta_\mathrm{DS}$-analysis picks out a potential substructure $320\arcsec$ ($\sim 1.0~\mathrm{Mpc}/h$) northeast of the BCG. In fact, four galaxies located at $\{ \theta_1, \theta_2\} \sim \{-200\arcsec, 250\arcsec \}$ ($\sim \{-0.64,  0.80\}~\mathrm{Mpc}/h$) have a chance in excess of $99.8\%$ to belong to a substructure. In conclusion, the $\Delta_\mathrm{DS}$ and the $\delta_\mathrm{DS}$-test bring further evidence for dynamical activity.

\section{Projection effects}
\label{app_proj}

The projected map $F_\mathrm{2D}$ of a volume density $F_\mathrm{3D}$ which is constant on surfaces of constant ellipsodial radius $\zeta$ is elliptical on the plane of the sky \citep{sta77,ser07},
\beq
\label{mult2}
F_\mathrm{2D} (\xi; l_\mathrm{P}, p_i) = \frac{2}{\sqrt{f}} \int_\xi^\infty F_\mathrm{3D}(\zeta; l_\mathrm{s}, p_i) \frac{\zeta}{\sqrt{ \zeta^2-\xi^2}} d \zeta,
\eeq
where $\xi$ is the elliptical radius in the plane of the sky, $l_\mathrm{s}$ is the typical length scale of the 3D density, $l_\mathrm{P}$ is its projection on the plane of the sky and $p_i$ are the other parameters describing the intrinsic density profile (slope, ...); the subscript P denotes measurable projected quantities. The parameter $f$ depends on the intrinsic shape and orientation of the 3D distribution,
\beq
\label{tri3}
f = e_1^2 \sin^2 \theta_\mathrm{Eu} \sin^2 \varphi_\mathrm{Eu} + e_2^2 \sin^2 \theta_\mathrm{Eu} \cos^2 \varphi_\mathrm{Eu}+ \cos^2 \theta_\mathrm{Eu},
\end{equation}
where $\varphi_\mathrm{Eu}$,  $ \theta_\mathrm{Eu}$ are the two Euler's angles of the principal cluster axes which fix the orientation of the line of sight and $e_i(\ge 1)$, $i=\{1,2\}$ are the two intrinsic axial ratios \citep{ser07}. The integral in Eq.~(\ref{mult2}) is proportional to $l_\mathrm{s}$. The relation between a length measured along the major axis and its projection in the sky is
\beq
\label{mult1}
\frac{l_\mathrm{s}}{\sqrt{f}} \equiv \frac{l_\mathrm{P}}{e_\Delta},
\eeq 
where the parameter $e_\Delta$ quantifies the elongation of the triaxial ellipsoid along the line of sight \citep{ser07},
\beq
e_\Delta = \left( \frac{e_\mathrm{P}}{e_1 e_2}\right)^{1/2} f^{3/4},
\eeq
with $e_\mathrm{P}(\ge 1)$ being the projected axial ratio. According to our notation for the NFW profile, the intrinsic $l_\mathrm{s}$ and the projected $l_\mathrm{P}$ have to be read as $r_\mathrm{s}$ and $r_\mathrm{sP}$,  respectively. Finally, the surface density can be expressed as
\beq
\label{mult3}
F_\mathrm{2D} = \frac{l_\mathrm{P}}{e_\Delta} f_\mathrm{2D} (\xi; l_\mathrm{P}, p_i,...) ,
\eeq 
where $f_\mathrm{2D}$ has the same functional form as for a spherically symmetric halo. Then, when we deproject a surface density, the normalization of the volume density can be known only apart from a geometrical factor 
\beq
f_\mathrm{geo} \equiv \frac{(e_1 e_2)^{1/2}}{f^{3/4}} = \frac{e_\mathrm{P}^{1/2}}{e_\Delta} .
\eeq 
When we estimate the gas mass from measurements of the surface brightness $S_\mathrm{X}$, which is proportional to the squared density $n_\mathrm{e}^2$, we first have to deproject $S_X$, so that, after inversion, the central density is known apart from a factor $f_\mathrm{geo}^{-1/2}$. Then we project along the line of sight the density $n_\mathrm{e}$, which brings about an additional factor $f_\mathrm{geo}$. The resulting central projected mass density is $\Sigma_0 \propto f_\mathrm{geo}^{1/2}$. This geometrical factor is independent of the specific density profile of the ICM distribution. 

When inferring the concentration parameter of the matter distribution, we face a slightly different case. We have just a single projection, so that the central convergence of a NFW profile estimated from lensing can be written in terms of $c_{200}$ and the projected length scale modulus a factor $f_\mathrm{geo}$,
\beq
\label{nfw1}
\Sigma_\mathrm{cr} \times \kappa_\mathrm{NFW} = \frac{f_\mathrm{geo}}{\sqrt{e_\mathrm{P}}}\rho_\mathrm{s} r_\mathrm{sP}.
\eeq
The estimate of the mass $M_{200}$ depends also on the scale-length $r_\mathrm{s}$ which is known modulus a factor $\sqrt{f}/e_\Delta$, see Eq.~(\ref{mult1}). Then
\beq
M_{200}= \frac{4\pi}{3}\times 200 \rho_\mathrm{cr} \times (c_{200} r_\mathrm{sP})^3 \frac{f_\mathrm{geo}}{e_\mathrm{P}^{3/2}},
\eeq
where $\rho_\mathrm{cr}$ is the critical density of the universe at the cluster redshift.

The geometrical correction factors can be estimated under some working hypotheses. Assuming that the cluster is drawn from a population with random orientations and intrinsic axial ratios with probability density following results from $N$-body simulations \citep{ji+su02}, we can estimate for $f_\mathrm{geo}$ ($f_\mathrm{geo}^{1/2}$) an expected value of 0.93 (0.95) with a dispersion of 0.37 (0.18). The concentration is obtained from Eq.~(\ref{nfw1}) after expressing $\rho_\mathrm{s}$ in terms of $c_{200}$ for an ellipsoidal halo \citep{co+ki07}.

%\bibliographystyle{aa}
%\bibliography{ac114_lensing}

\end{document}